\documentclass[fleqn,usenatbib]{mnras}

\usepackage{newtxtext,newtxmath}

\usepackage[T1]{fontenc}

\DeclareRobustCommand{\VAN}[3]{#2}
\let\VANthebibliography\thebibliography
\def\thebibliography{\DeclareRobustCommand{\VAN}[3]{##3}\VANthebibliography}

\usepackage{graphicx}	
\usepackage{amsmath}	
\usepackage{bm}
\usepackage[normalem]{ulem}

\newcommand{\msun}{{\,\rm M_\odot}}

\newcommand{\kms}{\,{\rm km}\,{\rm s}^{-1}}

\newcommand{\erg}{\,{\rm erg}}

\newcommand{\Myr}{\,{\rm Myr}}

\newcommand{\Mpc}{\,{\rm Mpc}}

\newcommand{\mmag}{\,{\rm mag}}

\def\jcap{J. Cosmol.  Astropart. Phys.}
\def\aap{A\&A}
\def\apj{ApJ}

\def\apjl{ApJ}
\def\mnras{MNRAS}
\def\araa{ARA\&A}
\def\aj{AJ}

\def\nat{Nature}
\def\natastro{Nature Astronomy}

\def\apjs{ApJS}

\title[]{The impact of UV variability on the abundance of bright galaxies at $z \geq 9$}

\author[Shen et al.]{\parbox{18.0cm}{
Xuejian Shen,$^{1,2}$\thanks{E-mail: xshen@caltech.edu}
Mark Vogelsberger,$^{2}$
Michael Boylan-Kolchin,$^{3}$
Sandro Tacchella,$^{4,5}$
and Rahul Kannan$^{6}$
}\vspace{0.3cm} \\
$^{1}$ TAPIR, California Institute of Technology, Pasadena, CA, 91125\\
$^{2}$ Department of Physics \& Kavli Institute for Astrophysics and Space Research, Massachusetts Institute of Technology, Cambridge, MA 02139, USA \\
$^{3}$ Department of Astronomy, The University of Texas at Austin, 2515 Speedway Stop C1400, Austin, TX 78712, USA \\ 
$^{4}$ Kavli Institute for Cosmology, University of Cambridge, Madingley Road, Cambridge, CB3 0HA, UK \\
$^{5}$ Cavendish Laboratory, University of Cambridge, 19 JJ Thomson Avenue, Cambridge, CB3 0HE, UK \\
$^{6}$ Department of Physics and Astronomy, York University, 4700 Keele Street, Toronto, ON M3J 1P3, Canada
}

\date{Accepted XXX. Received YYY; in original form ZZZ}

\pubyear{2023}

\begin{document}
\label{firstpage}
\pagerange{\pageref{firstpage}--\pageref{lastpage}}
\maketitle

\begin{abstract}
JWST observations have revealed a population of galaxies bright enough that potentially challenge standard galaxy formation models in the $\Lambda$CDM cosmology. Using a minimal empirical framework, we investigate the influence of variability on the rest-frame ultra-violet (UV) luminosity function (UVLF) of galaxies at $z\geq 9$. Our study differentiates between the \textit{median UV radiation yield} and the \textit{variability of UV luminosities} of galaxies at a fixed dark matter halo mass. We primarily focus on the latter effect, which depends on halo assembly and galaxy formation processes and can significantly increase the abundance of UV-bright galaxies due to the upscatter of galaxies in lower-mass haloes. We find that a relatively low level of variability, $\sigma_{\rm UV} \approx 0.75$ mag, matches the observational constraints at $z\approx 9$. However, increasingly larger $\sigma_{\rm UV}$ is necessary when moving to higher redshifts, reaching $\sigma_{\rm UV} \approx 2.0\,(2.5)\mmag$ at $z\approx 12$ ($16$). This implied variability is consistent with expectations of physical processes in high-redshift galaxies such as bursty star formation and dust clearance during strong feedback cycles. Photometric constraints from JWST at $z\gtrsim 9$ therefore can be reconciled with a standard $\Lambda$CDM-based galaxy formation model calibrated at lower redshifts without the need for adjustments to the median UV radiation yield.
\end{abstract}




\section{Introduction}

The James Webb Space Telescope (JWST) has opened a new window into the early and distant Universe, enabling studies of galaxy formation and evolution within the first $\sim 500\Myr$ ($z\gtrsim 10$) of the age of the Universe. Early JWST/NIRCam imaging datasets have led to the discovery of numerous photometric drop-out galaxy candidates at $z\gtrsim 9$~\citep[e.g.][]{Naidu2022,Castellano2022,Finkelstein2022,Adams2023,Atek2023,Bouwens2023a,Donnan2023,Harikane2023,Robertson2023,Yan2023} and even unusually bright galaxy candidates at $z\approx 16$~\citep{Donnan2023,Harikane2023}. 

The UV luminosities and estimated stellar masses of these sources have raised two key tensions. The first tension is related to the large stellar mass of some JWST-identified galaxies~\citep[e.g.][]{Labbe2023}, implying that the stellar mass density at $z\approx 7.5-9$ is comparable to the total mass budget of baryons within sufficiently massive dark matter haloes in a $\Lambda$CDM universe~\citep{BK2023,Lovell2023}. This result has been actively debated in the literature and is subject to many systematic uncertainties~\citep[e.g.][]{Endsley2022,Larson2022, Steinhardt2022, Chen2023, Prada2023}. The stellar masses and star-formation rates (SFRs) of individual galaxies at $z \gtrsim 10$ thus far appear consistent with the standard structure formation theory~\citep[e.g.][]{Keller2023, McCaffrey2023}. 

The second tension concerns the abundance of UV bright galaxies, which is more robust and will be the focus of this paper. Although characteristic shapes of the rest-frame UV luminosity functions (UVLFs) determined using JWST-identified galaxies are consistent with those derived with the Hubble Space Telescope (HST) observations, the bright end of the UVLFs shows little evolution beyond $z\approx 10$ and lacks the steep decline expected from extrapolating Schechter function fits from lower redshifts~\citep[e.g.][]{Harikane2023,Finkelstein2023}. As a result, the implied star-formation rate density (SFRD) declines only slowly at $z\gtrsim 10$, in contrast to the rapid decline predicted by constant star-formation efficiency models~\citep[e.g.][]{Bouwens2023b,Harikane2023}. 

Even after accounting for various observational corrections \citep[e.g.][]{Finkelstein2023}, the suggested abundance of UV bright galaxies at $z\gtrsim 10$ surpasses theoretical predictions from a wide range of models. This includes empirical models~\citep[e.g.][]{Tacchella2013,Mason2015,Sun2016,Tacchella2018,Behroozi2020}, semi-analytical galaxy formation models~\citep[e.g.][]{Dayal2014,Dayal2019,Yung2019,Mauerhofer2023,Yung2023}, and cosmological hydrodynamic simulations~\citep[e.g.][]{Simba,Vogelsberger2020,Haslbauer2022,Kannan2022-thesan,Kannan2022,Wilkins2023,Wilkins2023b} that have been calibrated for lower redshift galaxies. One possible explanation for this discrepancy is that the early results are based on the photometrically-selected galaxy candidates, which may be contaminated by low-redshift interlopers \citep[e.g.][]{Zavala2023, Naidu2022b, Fujimoto2022}. However, recent pure spectroscopic constraints of the UVLF \citep[e.g.][]{AH2023b,AH2023a,Curtis2023,Robertson2023,Harikane2023-spec} yield broadly consistent results with the photometric estimates. 

The UVLF tension between observations and predictions suggests that our current understanding of galaxy formation in the early Universe may need to be revised. Several physical interpretations have been discussed in the literature to explain the tension~\citep[e.g.][]{Inayoshi2022,Ferrara2022,Dekel2023,Mason2023,Yung2023}. These include but are not limited to (1) a substantially higher star-formation efficiency for normal stellar populations, (2) a top-heavy stellar initial mass function (IMF), (3) zero dust attenuation, (4) UV radiation contributed by non-stellar sources, e.g. accreting stellar-mass black holes, quasars/active galactic nuclei. These solutions primarily aim to enhance the \textit{median} UV radiation yield from early galaxies. Another direction suggested by e.g. \citet{Mason2023} and \citet{Mirocha2023} involves increased stochasticity of star formation such that galaxies in a temporary high-SFR phase will appear as the UV luminous sources. The steep decline of the underlying halo mass function in the massive/bright end means that there are more intrinsically low-mass sources upscattered to high luminosities than massive galaxies downscattered to faint luminosities, which will populate the bright end of the UVLF. 

In this paper, we examine the UV variability of high-redshift galaxies coming from a variety of sources of stochasticity, including halo assembly, star formation, and dust attenuation, in the context of a canonical \citet{Salpeter1955} IMF. We investigate the impact of UV variability on galaxy UVLFs, focusing on the constraints imposed by JWST observations at $z\geq 9$. We will study this using an empirical approach and decompose the effects of variability and the shift of the median galaxy-halo connection. The paper is organized as follows: In Section~\ref{sec:method}, we first introduce the model establishing a median mapping between the halo mass function and galaxy UVLF. We then describe how we treat UV variability. In Section~\ref{sec:results}, we present the results and discuss the implication of UV variability in reconciling JWST observations with a standard galaxy formation model in $\Lambda$CDM. In Section~\ref{sec:conclusions}, we provide our conclusions.

\section{Method}
\label{sec:method}

\subsection{Median galaxy UV luminosity}
\label{sec:method:median}

We adopt the flat $\Lambda$CDM cosmological model of \citet{Planck2020}, assuming the primordial density fluctuations are Gaussian and adiabatic. The  cosmological parameters relevant for this study are $h \equiv H_0/(100\,\kms \Mpc^{-1}) = 0.6732$, $\Omega_{\rm m} = 0.3158$, $n_{\rm s} = 0.96605$, $\sigma_{8} = 0.8120$, and $f_{\rm b} \equiv \Omega_{\rm b}/\Omega_{\rm m} = 0.156$. \\[0.2cm]
\noindent\textbf{Halo mass function:} The halo mass function is constructed following Press-Schechter-like theories \citep[e.g.][]{Press1974, Bond1991, Sheth2001} as implemented in the {\sc Hmf} code~\citep{hmf2,hmf3}. We adopt the transfer function calculated using the Code for Anisotropies in the Microwave Background ({\sc Camb}; \citealt{CAMB1,CAMB2}), the halo mass function parametrization of \citet{Tinker2010}, 
and a real-space top-hat filter function for the density field. The definition of halo mass follows the virial criterion in \citet{Bryan1998}. \\[0.2cm]
\noindent\textbf{Halo accretion rate:} We use the fitting function of \textit{median} halo accretion rate in \citet{Fakhouri2010}
\begin{align}
    \dot{M}_{\rm halo} (M_{\rm halo}, z) & \simeq 25.3 \msun\,{\rm yr}^{-1}\,\left( \dfrac{M_{\rm halo}}{10^{12}\msun} \right)^{1.1} \nonumber \\
    & \times (1+1.65\,z)\, \sqrt{\Omega_{\rm m}\,(1+z)^{3} + \Omega_{\Lambda}},
\end{align}
which is calibrated on the joint data set from the Millennium and Millennium-{\small II} simulations~\citep{Springel2005,BK2009}~\footnote{The cosmological parameters adopted in these simulations are out of date. As discussed in \citet{Inayoshi2022}, the impact on the halo growth rate is limited ($\lesssim 0.1\,{\rm dex}$), as found in \citet{Dong2022} using up-to-date cosmological parameter sets.}. \\[0.2cm]
\noindent\textbf{Star formation:} We parameterize the SFR in dark matter haloes as ${\rm SFR} = \varepsilon_{\ast}\,f_{\rm b}\,\dot{M}_{\rm halo}$, where $f_{\rm b}$ is the universal baryon fraction and $\varepsilon_{\ast}$ is the star-formation efficiency. We adopt a \textit{redshift-independent} double power-law function, 
\begin{equation}
    \varepsilon_{\ast}(M_{\rm halo}) = \dfrac{2 \, \varepsilon_{0}}{ (M_{\rm halo}/M_{0})^{-\alpha} + (M_{\rm halo}/M_{0})^{\beta} }\,,
    \label{eq:sfe}
\end{equation}
where $\varepsilon_{0}$ is the peak star-formation efficiency at the characteristic mass $M_{0}$ and $\alpha$ and $\beta$ are the low-mass and high-mass end slopes, respectively. The functional form and the redshift-independent ansatz of Equation~\ref{eq:sfe} have been used in previous empirical modeling works~\citep[e.g.][]{Moster2010,Tacchella2018,Harikane2022}. We adopt $\varepsilon_{0}=0.1$, $M_{0}=10^{12}\msun$, $\alpha=0.6$, $\beta=0.5$ as our default values. The normalization and low-mass slope are chosen to match the median SFR-$M_{\rm halo}$ relation at $z=7$ from \citet{Behroozi2019} while the high-mass slope follows the value in \citet{Harikane2022}. The parameter choices give good agreement with the observed UVLFs and UV luminosity densities at $z\lesssim 9$. At the halo mass scale that is typical for bright JWST-detected galaxies ($M_{\rm halo} \sim 10^{10}\msun$), $\varepsilon_{\ast}$ takes the value of $\sim 0.01$. This model is a basic representation~\footnote{Star formation in high-redshift galaxies potentially exhibits complex dependencies on numerous factors that are not captured by a straightforward empirical model. There are substantial uncertainties in constraining these factors with available observational data along with strong model dependence in the calibration procedure. Acknowledging these challenges, the main goal of this empirical model is to create a rough reference point, representing models built prior to the introduction of JWST data. Subsequently, we aim to parameterize model variations by the shift in the median galaxy-halo connection with respect to this reference model and variability.} of our knowledge about galaxy formation prior to the JWST era. \\[0.2cm]

\noindent\textbf{SFR -- UV luminosity:} We express the conversion between the SFR and the intrinsic UV-specific luminosity $L_{\nu}(\mathrm{UV})$
(before dust attenuation) as
\begin{equation}
    \label{eq:sfr-luv}
    \mathrm{SFR}\,[\msun\,{\rm yr}^{-1}] = \kappa_{\rm UV} \, L_{\nu}(\mathrm{UV})\,[\erg\,{\rm s}^{-1}\,{\rm Hz}^{-1}]
\end{equation}
with conversion factor $\kappa_{\rm UV} = 1.15 \times 10^{-28}$ as in \citet{Madau2014}, where a \citet{Salpeter1955} IMF is assumed and the (far-)UV wavelength is assumed to be $1500$\AA. \\[0.2cm]
\noindent\textbf{Dust attenuation:} We empirically model dust attenuation using a combination of the $A_{\rm UV}$-$\beta$ (IRX-$\beta$) relation and $\beta$-$M_{\rm UV}$ relation. We adopt the relation $A_{\rm UV} = 4.43 + 1.99 \,\beta$ \citep{Meurer1999} and the most recent $\beta$-$M_{\rm UV}$ relation $\beta = -0.17\, M_{\rm UV} - 5.40$ at $z\gtrsim 8$ from \citet{Cullen2023}. Combine the two relations, we obtain a \textit{median} attenuation at a given $M_{\rm UV}$ of
\begin{equation}
    \label{eq:dust}
    A_{\rm UV} = - 0.34\, \left[21+M_{\rm UV}\right] + 0.79\,.
\end{equation}
The $M_{\rm UV}$ here is the observed (dust-attenuated) UV magnitude. The recipe gives median $A_{\rm UV}= [0.45, 0.79, 1.13]\mmag$ attenuation at observed $M_{\rm UV} = [-20, -21, -22] \mmag$.

\begin{figure*}
    \centering
    \includegraphics[width=0.495\linewidth]{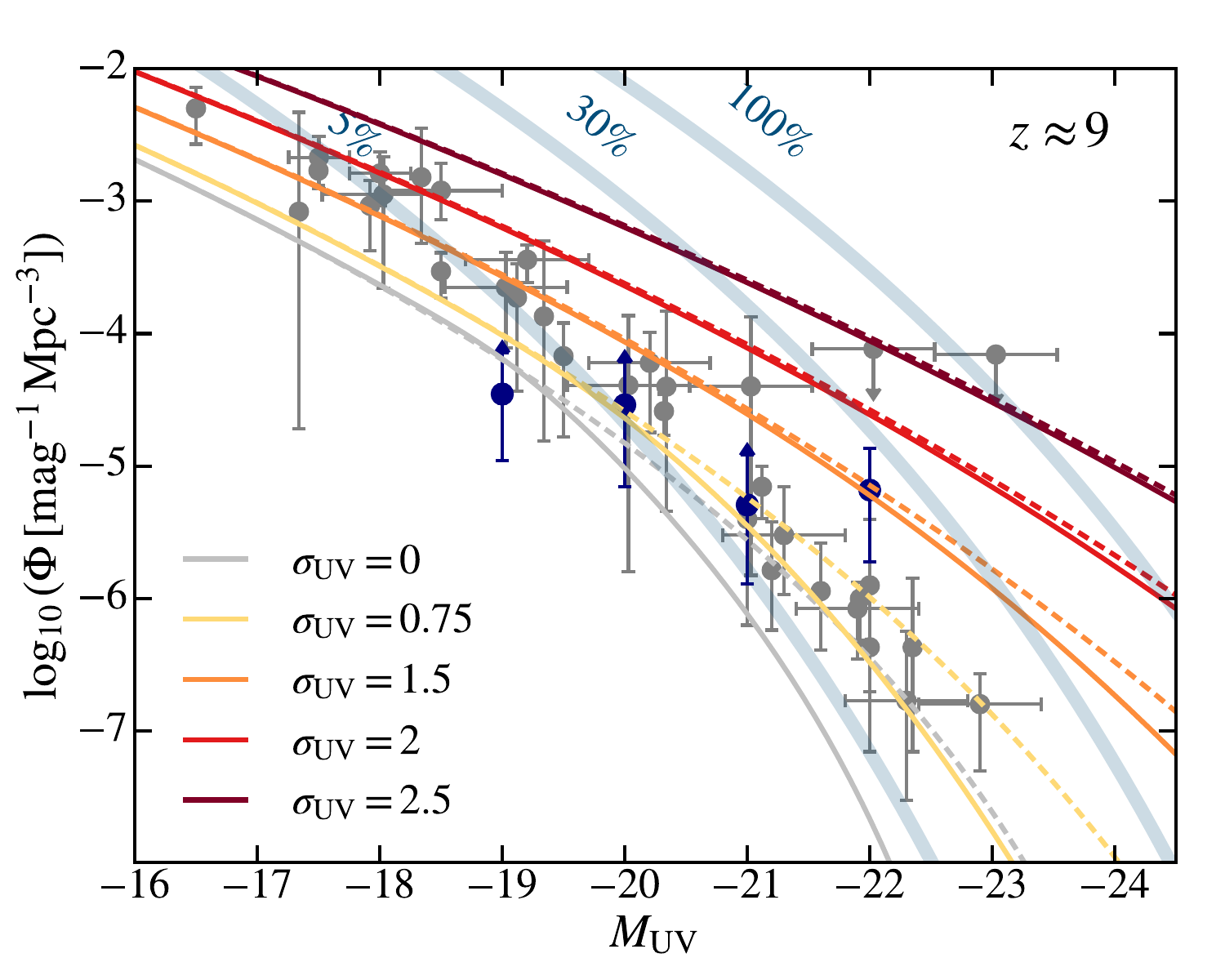}
    \includegraphics[width=0.495\linewidth]{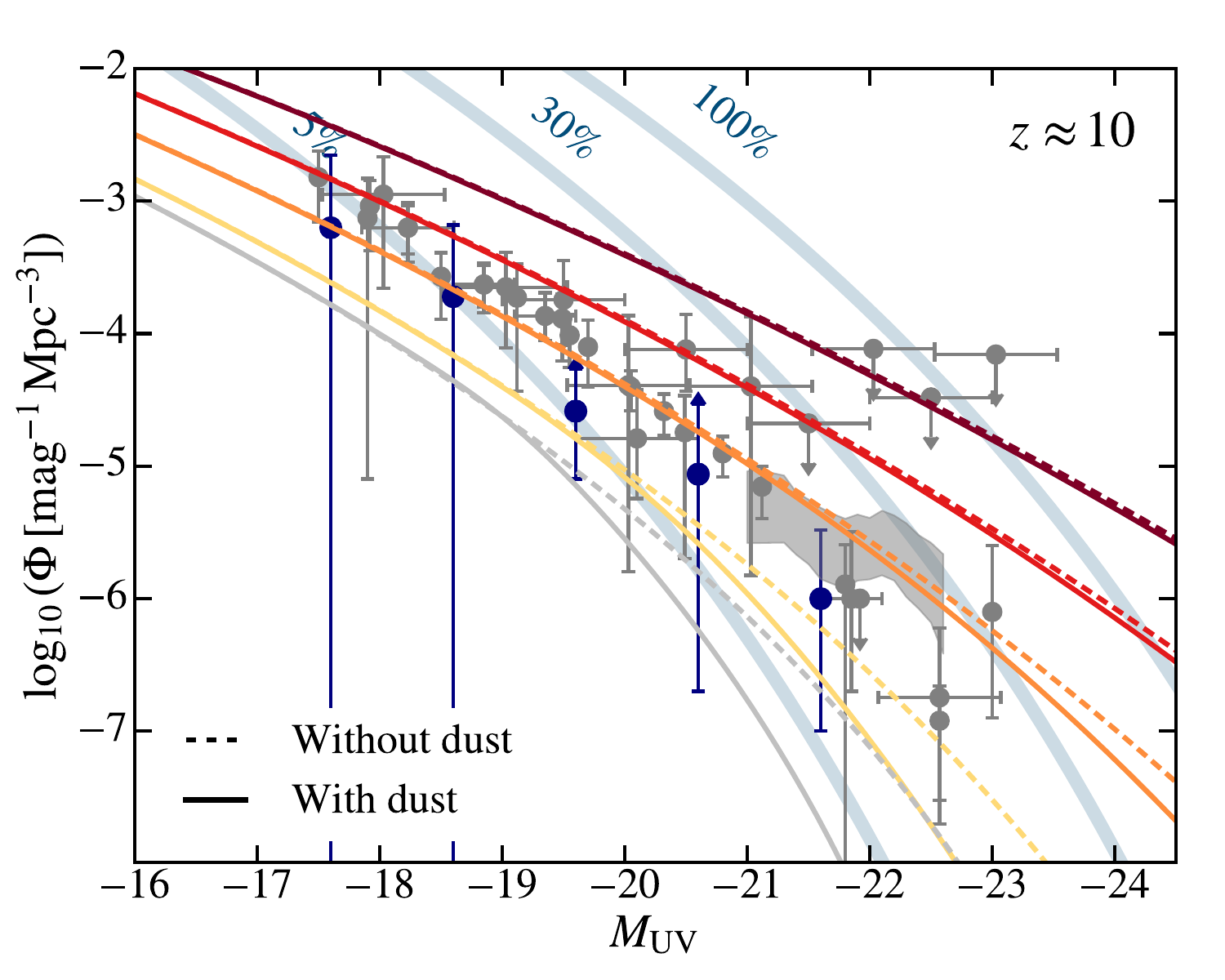}
    \includegraphics[width=0.495\linewidth]{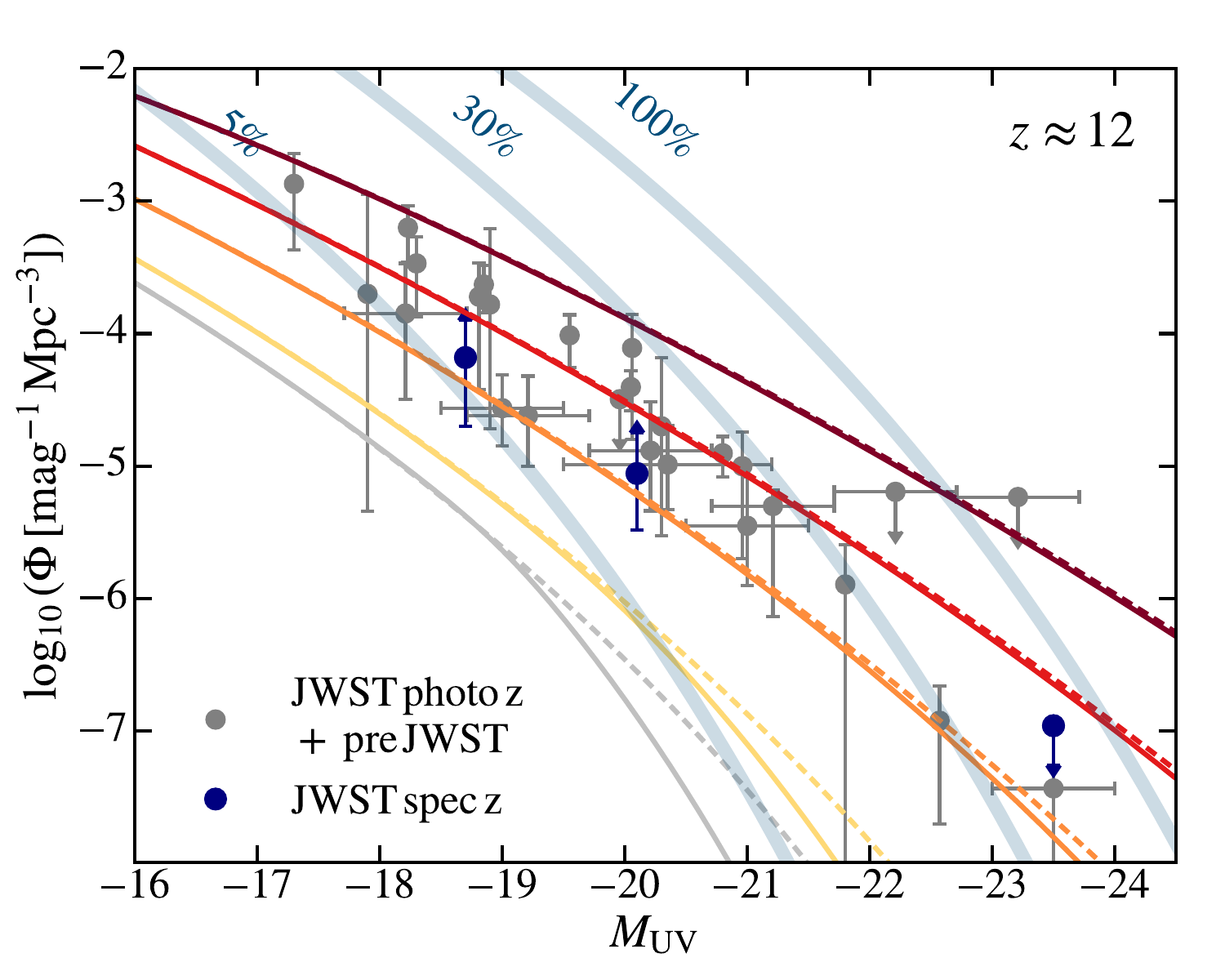}
    \includegraphics[width=0.495\linewidth]{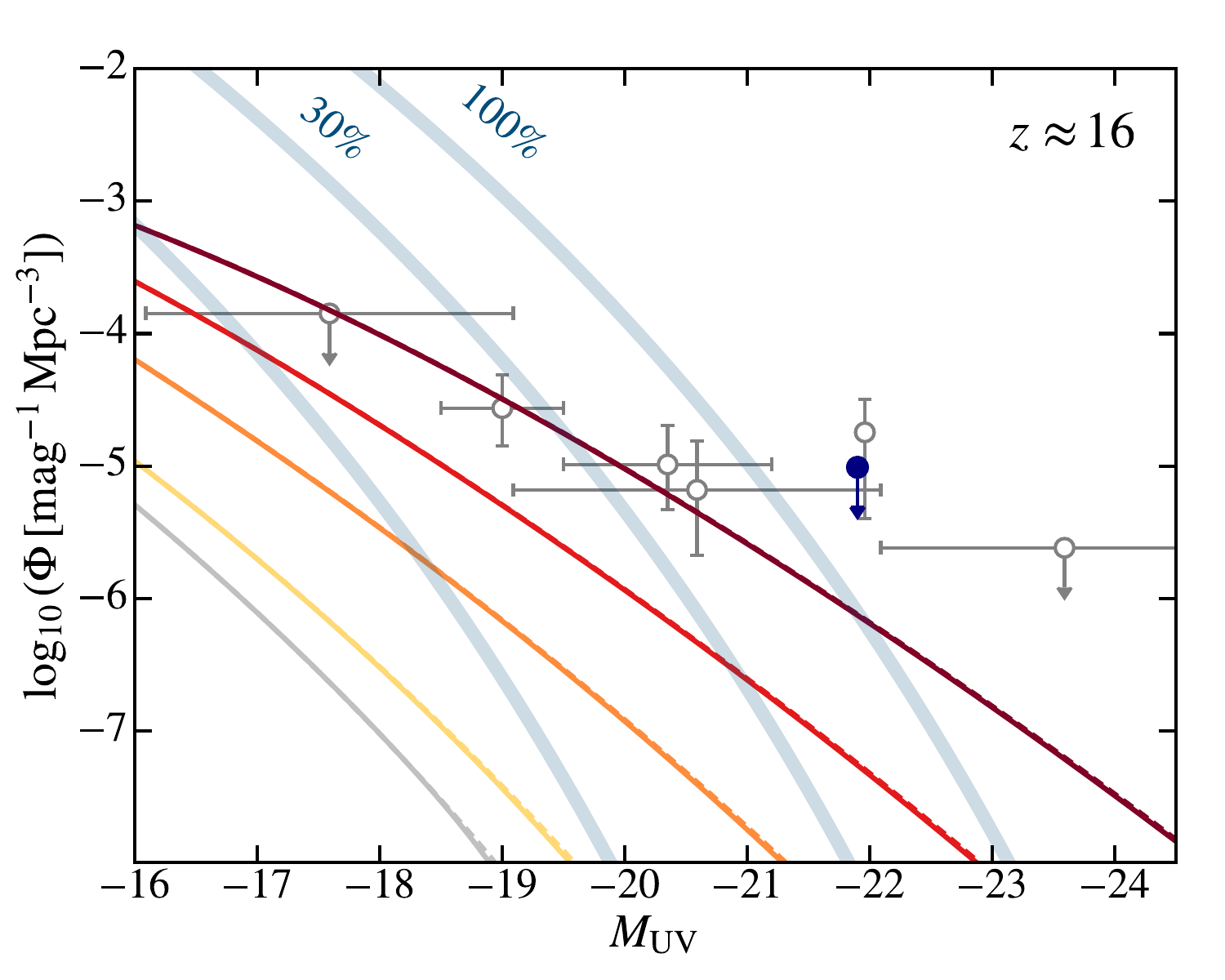}
    \caption{Rest-frame UV luminosity functions (UVLFs) of galaxies at $z\gtrsim 9$. The blue lines represent UVLFs assuming a constant star-formation efficiency $\varepsilon_{\ast}$ of the value marked. The gray data points are measurements based on photometrically-selected galaxies by JWST~\citep{Castellano2022, Finkelstein2022, Naidu2022, Adams2023b, Bouwens2023a, Bouwens2023b, Donnan2023, Harikane2023, Leetho2023, McLeod2023, Morishita2023, Perez2023} as well as pre-JWST constraints~\citep{McLeod2016,Oesch2018,Morishita2018,Stefanon2019,Bowler2020,Bouwens2021}. The photometric constraints at $z\approx 16$ are highly uncertain and therefore shown with open markers. The dark blue data points are based on the JWST spectroscopically-confirmed galaxies (e.g.~\citealt{AH2023b,AH2023a,Bunker2023,Curtis2023}; see the full references in \citealt{Harikane2023-spec}). Assuming our default halo-mass-dependent $\varepsilon_{\ast}$ (Equation~\ref{eq:sfe}), UV variability of $\sigma_{\rm UV} \approx 1.5$, $2.0$, $2.5$ mag is required to match the JWST photometric constraints at $z\approx 10$, $12$, $16$. At $z \approx 9$, a lower value of $0.75$ mag is preferred to match observations at the bright end. 
    }
    \label{fig:uvlf}
\end{figure*}

\subsection{UV variability}
\label{sec:method:sigmaUV}

The model above describes a \textit{median} mapping from the dark matter halo mass to the observed galaxy UV luminosity. It allows us to calculate galaxy rest-frame UVLF based on the underlying halo mass function
\begin{equation}
    \dfrac{{\rm d}n}{{\rm d}M_{\rm UV}} = \dfrac{{\rm d}n}{{\rm d} \log_{10}{M_{\rm halo}}} \left| \dfrac{{\rm d} \log_{10}{M_{\rm halo}}}{{\rm d}M_{\rm UV}} \right|.
\end{equation}
However, stochasticity in both halo assembly and galaxy formation processes can give rise to scatter with respect to the median $M_{\rm UV}-M_{\rm halo}$ relation. This manifests as the scatter in all scaling relations we introduced above. To model this stochasticity, we convolve the UV luminosity function with a Gaussian kernel~\footnote{The choice of the convolution kernel here is motivated by the observed or theoretically-predicted distribution of halo accretion rates~\citep[e.g.][]{Fakhouri2010,Ren2019,Mirocha2021} and star formation rates of high-redshift galaxies~\citep[e.g.][]{Speagle2014,Popesso2023,Pallottini2023}, which contribute to the UV variability. The same kernel is adopted in many previous studies that include the variability effects~\citep[e.g.][]{Ren2018,Whitler2020}.} of width $\sigma_{\rm UV}$ (in unit of AB magnitude). Effectively, this assumes that the observed UV luminosity has a log-normal distribution with the \textit{median} value fixed. Note that this will increase the \textit{mean} UV luminosity by a factor of
\begin{equation}
    \dfrac{ \langle L^{\rm conv}_{\nu}({\rm UV}) \rangle}{\langle L_{\nu}({\rm UV}) \rangle} = \exp{\left( \dfrac{ (\ln{10}\,\sigma_{\rm UV}/2.5)^2}{2} \right)},
\end{equation}
which is equivalent to roughly $0.5\,\sigma^2_{\rm UV}$ mag. Moreover, due to the steeply decreasing nature of the halo mass function and the UVLF, the upscatter in UV luminosity dominates over the downscatter, leading to an enhanced abundance of UV luminous galaxies. This effect will be demonstrated and discussed in the results section of the paper. We use the term ``UV variability'' to summarize this scatter in the $M_{\rm UV}$-$M_{\rm halo}$ relation since the variations of luminosities of single sources over short time scales could contribute significantly to the scatter. The potential source of UV variability include:\\[0.2cm]
\noindent\textbf{Halo assembly}: The mass accretion rate of dark matter haloes roughly follows a log-normal distribution with a typical $1\,\sigma$ scatter of $\sigma_{\rm halo} \approx 0.3\,{\rm dex}$, which is broadly consistent with cosmological $N$-body simulations \citep[e.g.][]{Rodrguez2016, Ren2019, Mirocha2021, Mirocha2023}. It is independent of any baryonic processes.\\[0.2cm]
\noindent\textbf{Star formation}: In both simulations and observations, small dwarf galaxies and high-redshift galaxies exhibit ``bursty'' star-formation histories~\citep[e.g.][]{Sparre2015,Smit2016,Emami2019,Iyer2020,Tacchella2020,Flores2021,Hopkins2023} characterized by large temporal fluctuations in SFR. This bursty phase aligns with the irregular and clumpy morphologies of the observed high-redshift galaxies~\citep[e.g.][]{Bournaud2007,Elmegreen2009,Forster2011,Treu2023}. Large scatter in star-formation efficiency can be driven by the interplay of gas inflow/outflow, instability, and galaxy mergers in the early phase of galaxy formation \citep[e.g.][]{Dekel2009,Ceverino2010,Angles2017}, cycles of starbursts ceased by strong feedback~\citep[e.g.][]{ElBadry2016,Tacchella2016}, and some extreme feedback-free starbursts \citep[e.g.][]{FG2018,Dekel2023}. Galaxies in this phase are qualitatively different from sufficiently massive low-redshift star-forming galaxies in equilibrium stages, which exhibit smooth galaxy-integrated SFRs. The expected SFR variability in high-redshift galaxies is highly uncertain. The lower limit should be the scatter in star-formation efficiency ($\approx 0.15\,{\rm dex}$) inferred from observations at $z\lesssim 7$~\citep[e.g.][]{Harikane2018,Harikane2022}, and the main sequence scatter at high redshift, $\approx 0.3\,{\rm dex}$~\citep{Speagle2014}. We conservatively assume $\sigma_{\rm SF} \geq 0.3\,{\rm dex}$.\\[0.2cm]
\noindent\textbf{Dust attenuation}: Given the irregular and clumpy nature of high-redshift galaxies, the sightline and geometrical variations of dust attenuation can be large~\citep[e.g.][]{Carniani2018,Cochrane2019,Ferrara2022b}. In addition, the strong supernovae and radiative feedback, both temporally and spatially associated with the burst phase of star formation, can expel the majority of the cold phase gas and cause galaxies to temporarily become transparent to dust attenuation \citep[e.g.][]{Ferrara2022,Flore2023,Nath2023,Ziparo2023}. The degree of UV variability contributed or balanced off by these factors depends on the amount of dust in these galaxies as well as the coherence between the dust clearance and the starburst. The scatter in the observed $\beta$-$M_{\rm UV}$ relation is found to be $\sigma_{\beta} \approx 0.35$~\citep{Bouwens2014,Rogers2014,Cullen2023}, which corresponds to $\sigma_{\rm dust} \approx 0.7\mmag$ for our assumed $A_{\rm UV}$-$\beta$ relation. \\[0.2cm]
\noindent\textbf{Bracketing the combined effect:} The true UV variability in high-redshift galaxies, as well as its potential dependence on halo mass or redshift, are challenging to constrain given the uncertainties in the physical drivers, the limited observational probes, and errors. Consequently, we maintain it as a free parameter throughout our analysis while adhering to specific constraints. To model the combined effects of the three sources of variability mentioned above, we numerically sample haloes based on the halo mass function and calculate their observed UV magnitudes individually. We model the halo mass accretion rate, $\dot{M}_{\rm halo}$, and the star formation efficiency, $\varepsilon_{\ast}$, as log-normal distributions, while the dust attenuation, $A_{\rm UV}$, is modeled as a normal distribution. Their median values are determined as in Section~\ref{sec:method:median}. The $1\,\sigma$ scatters are $\sigma_{\rm halo}$, $\sigma_{\rm SF}$, and $\sigma_{\rm dust}$, as estimated above, and are assumed to be independent of $M_{\rm halo}$ and $z$. Owing to the influence of dust attenuation (Equation~\ref{eq:dust}), the relationship between $M_{\rm UV}$ and $\log_{10}{M_{\rm halo}}$ is non-linear. The distribution of observed UV luminosities for galaxies at a fixed halo mass, therefore, does not strictly follow a log-normal distribution. To define the effective $\sigma_{\rm UV}$, we match the numerically sampled UVLF with the one obtained through convolution using Gaussian kernels of width $\sigma_{\rm UV}$ at $M_{\rm UV} \approx -21$ at $z\approx 10$ (our results are insensitive to these assumed values of $M_{\rm UV}$ and $z$).  We consider three typical cases. (1) If we account for only $\sigma_{\rm halo}$ while ignoring scatter in $\varepsilon_{\ast}$ and $A_{\rm UV}$, we obtain $\sigma_{\rm UV} \approx 0.6\,(0.75)\mmag$ with (without) dust attenuation, which sets the minimum UV variability. (2) If the scatters in $\dot{M}_{\rm halo}$, $\varepsilon_{\ast}$, and $A_{\rm UV}$ are perfectly correlated, we obtain $\sigma_{\rm UV} \gtrsim 2.2\mmag$, which represents the maximum UV variability. (3) If the scatters in $\dot{M}_{\rm halo}$, $\varepsilon_{\ast}$, and $A_{\rm UV}$ are independent, we obtain $\sigma_{\rm UV}\gtrsim 1.2\mmag$, which serves as a more conservative estimate.

\section{Results}
\label{sec:results}

In Figure~\ref{fig:uvlf}, we present the UVLF calculated at $z \gtrsim 9$ assuming different levels of UV variability. For comparison, we show the observational constraints based on the photometrically-selected JWST sources~\citep{Castellano2022, Finkelstein2022, Naidu2022, Adams2023b, Bouwens2023a, Bouwens2023b, Donnan2023, Harikane2023, Leetho2023, McLeod2023, Morishita2023, Perez2023}, pre-JWST constraints~\citep{McLeod2016,Oesch2018,Morishita2018,Stefanon2019,Bowler2020,Bouwens2021}, and the constraints based on pure spectroscopically-confirmed samples compiled in \citet{Harikane2023-spec}. We note that the $z\approx 16$ constraints are based on a few photometrically-selected galaxy candidates and therefore highly uncertain. For example, one previously-claimed $z\approx 16$ galaxy candidate first identified in \citet{Donnan2023} was found to be a galaxy at $z = 4.912$~\citep{AH2023a}. In addition, at $z\approx 10, 12$, the photometric redshifts of galaxy candidates have non-negligible scatter~\citep[e.g.][]{Finkelstein2023,McLeod2023}. A safer way to approach the problem is to consider the full redshift probability distribution for each galaxy candidate when constructing the UVLF, but it is beyond the scope of this study.

Models with a constant star-formation efficiency and zero UV variability require $\varepsilon_{\ast} \gtrsim 30\%$ to explain the most stringent observational results at $z\geq 10$, which is much higher than the canonical value $\varepsilon_{\ast} \lesssim 5\%$ for $z\lesssim 9$ galaxies in a similar mass range. Such models also fail to reproduce the shape of observed UVLFs: they have steeper faint-end slopes and more abrupt exponential cutoffs than observations. Adopting the halo mass-dependent star-formation efficiency $\varepsilon_{\ast}(M_{\rm halo})$ from Equation~\ref{eq:sfe} helps make the shape of the UVLF more consistent with observations. Nevertheless, in the absence of UV variability, the model systematically underpredicts the luminosity of galaxies. Using $\sigma_{\rm UV} = 0.75$ mag, our assumed minimum value (coming solely from scatter in halo accretion rates), leads to a UVLF that is consistent with the $z=9$ observational results at the bright end. Similar value has been found in previous studies~\citep[e.g.][]{Ren2018,Ren2019,Whitler2020}. At the faint end at $z=9$, a larger $\sigma_{\rm UV} \approx 1.5\mmag$ is required. This is a general trend at all redshifts studied, which is consistent with low-mass galaxies having more bursty star formation. However, we caution that any mass or luminosity dependence of $\sigma_{\rm UV}$ may be degenerate with assumptions about the explicit halo mass dependence of $\varepsilon_{\ast}$ (i.e., smaller values of $\alpha$ in Equation~\ref{eq:sfe}).

As the UV variability increases, the abundance of luminous galaxies is enhanced. $\sigma_{\rm UV} = 1.5$, $2.0$, $2.5\mmag$ is sufficient to explain current JWST constraints at $z = 10$, $12$, $16$, even assuming \textit{all} photometrically-selected candidates are real. This level of UV variability can be contributed by additional variances in star-formation efficiency and dust attenuation, with potentially large correlations with the variation of halo accretion rates (as discussed in Section~\ref{sec:method:sigmaUV}). Similar values tend to overproduce galaxies at $z \leq 9$, indicating a qualitative transition in UV variability at $z\approx 10$. These UV-bright phase galaxies are expected to have very blue intrinsic colors in UV \citep[e.g.][]{Topping2022,Adams2023,Atek2023,Cullen2023} but could be balanced by dust attenuation \citep{Mirocha2023}, depending on how aligned the dust and the star-formation duty-cycles are. Considering the small fields probed by JWST, cosmic variance due to large-scale galaxy clustering could be significant. We refer to the estimates in \citet{Yung2023} using the online calculator of \citet{Trenti2008}. For typical effective survey areas of JWST ($\approx 10-35\,{\rm arcmin}^{2}$) at $z\gtrsim 10$, the cosmic variance is $\lesssim 0.2\,{\rm dex}$ in number density, which is subdominant compared to other observational uncertainties. Increasing UV variability will further decrease cosmic variance since the observed galaxies will correspond to lower-mass haloes, which are less clustered. 

\begin{figure}
    \centering
    \includegraphics[width=1\linewidth]{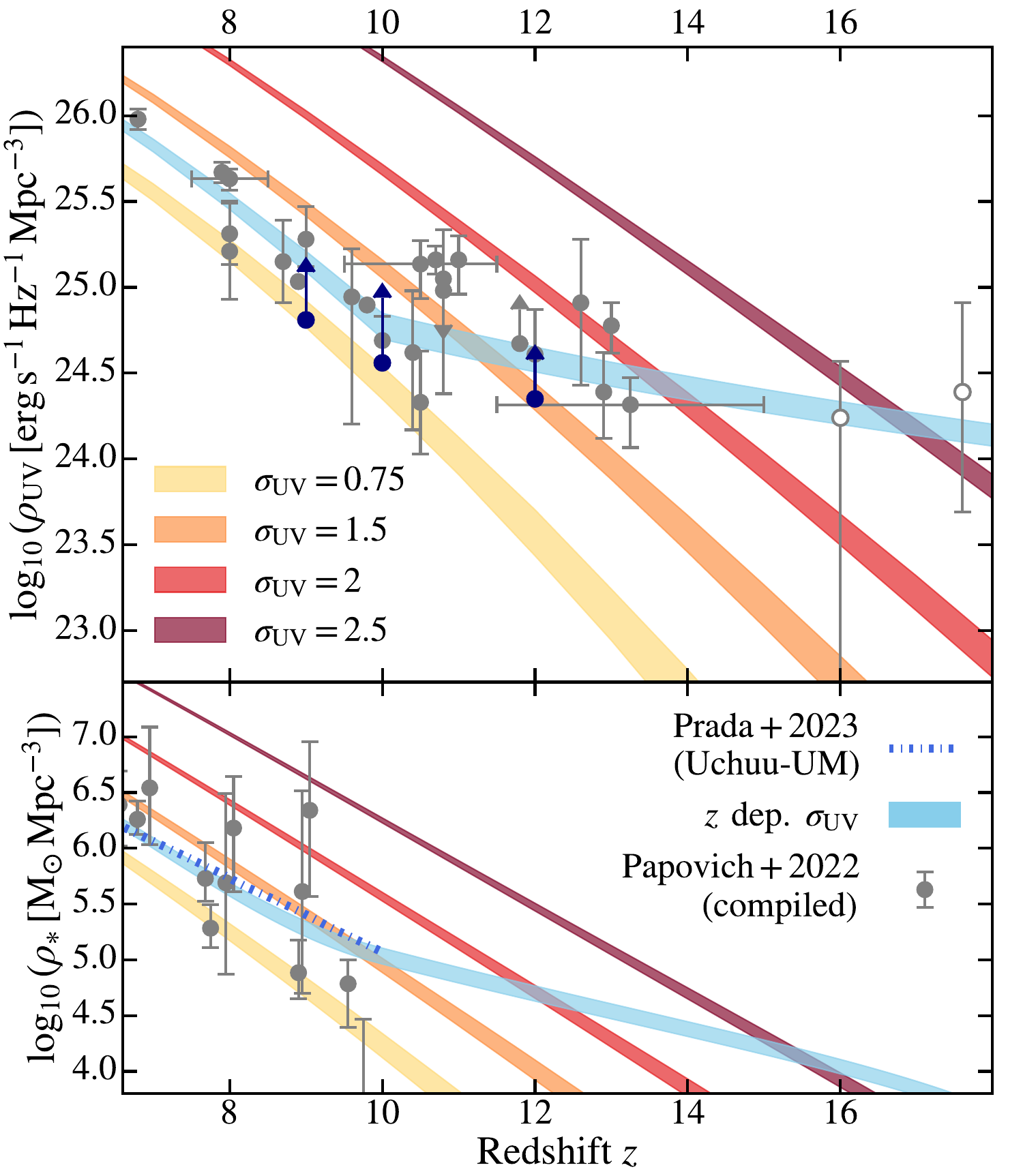}
    \caption{\textit{Top panel:} UV luminosity density $\rho_{\rm UV}$ of galaxies, integrated down to $M_{\rm UV}=-18$ (lower boundary of each shaded region) and $M_{\rm UV}=-17$ (upper boundary), as a function of redshift. The gray data points show photometric constraints~\citep{Coe2013,Ellis2013,Bouwens2020,Bouwens2023a,Bouwens2023b,Donnan2023,McLeod2023,Harikane2023,Perez2023} while the dark blue ones highlight pure spectroscopic constraints from \citet{Harikane2023-spec}. At $z\lesssim 10$, a relatively low UV variability ---  comparable to what is expected solely from variance in halo accretion rates --- is sufficient to explain the observations. At $z\gtrsim 10$, an increasingly large UV variability is required to explain the observational results. The cyan line shows a redshift-dependent $\sigma_{\rm UV}$ inferred from our comparison with JWST UVLFs (declining from $\sigma_{\rm UV} \approx 2.5\mmag$ at $z \approx 16$ to $\approx 0.75 - 1.5\mmag$ at $z \lesssim 10$). \textit{Bottom panel}: The cosmic stellar mass density obtained by integrating the SFR density from $z=20$, assuming the same limiting magnitude range. The results are compared with the latest observational constraints compiled in \citet{Papovich2023} and the predictions from the \textsc{Uchuu-um} model~\citep{Prada2023}. The cyan line shows the results assuming the redshift-dependent $\sigma_{\rm UV}$. The large $\sigma_{\rm UV}$ at early times does not lead to any discrepancies with the stellar mass density constraints at $z \lesssim 10$.}
    \label{fig:rhouv-redshift}
\end{figure}

The UV variability has a stronger influence on the bright end of the UVLF. However, integrated down to a canonical faint-end limit $M_{\rm UV} \approx -18$ to $-17$~\citep[e.g.][]{Bouwens2015, McLeod2016, Oesch2018, Bouwens2020, Harikane2023}, it still has a substantial impact. In the top panel of Figure~\ref{fig:rhouv-redshift}, we show the UV luminosity density integrated down to $M_{\rm UV}=-18$ (lower boundary of each shaded region) and $M_{\rm UV}=-17$ (upper boundary) as a function of redshift. They are compared to observational constraints~\citep{Coe2013,Ellis2013,Bouwens2020,Bouwens2023a,Bouwens2023b,Donnan2023,McLeod2023,Harikane2023,Perez2023}. The measurements based on spectroscopically-confirmed samples from \citet{Harikane2023-spec} are shown in blue. A low value of $\sigma_{\rm UV}$ between $0.75$ and $1.5\mmag$ works reasonably well in explaining the $\rho_{\rm UV}$ (and similarly for SFR density) at $z\lesssim 10$. A clear transition happens at $z\gtrsim 10$, where a larger UV variability $\sigma_{\rm UV} \gtrsim 1.5 \mmag$ is necessary to explain observational results if one maintains the same median galaxy-halo connection. In the bottom panel of Figure~\ref{fig:rhouv-redshift}, we show the cosmic stellar mass density by integrating the SFR density from $z=20$. The SFR density is converted from the UV luminosity density using Equation~\ref{eq:sfr-luv}, assuming the same limiting magnitude range. We highlight a redshift-dependent $\sigma_{\rm UV}$ scenario, where a large $\sigma_{\rm UV} \approx 2.5\mmag$ at $z \approx 16$ declines to $\approx 0.75 - 1.5\mmag$ at $z \approx 10$. We compared these results with the latest observational constraints compiled in \citet{Papovich2023} and the predictions from the \textsc{Uchuu-um} model~\citep{Prada2023}. We find the large UV variability at high redshift does not lead to any discrepancies with the stellar mass density constraints at $z \lesssim 10$. 

\begin{figure}
    \raggedright
    \includegraphics[width=1\linewidth]{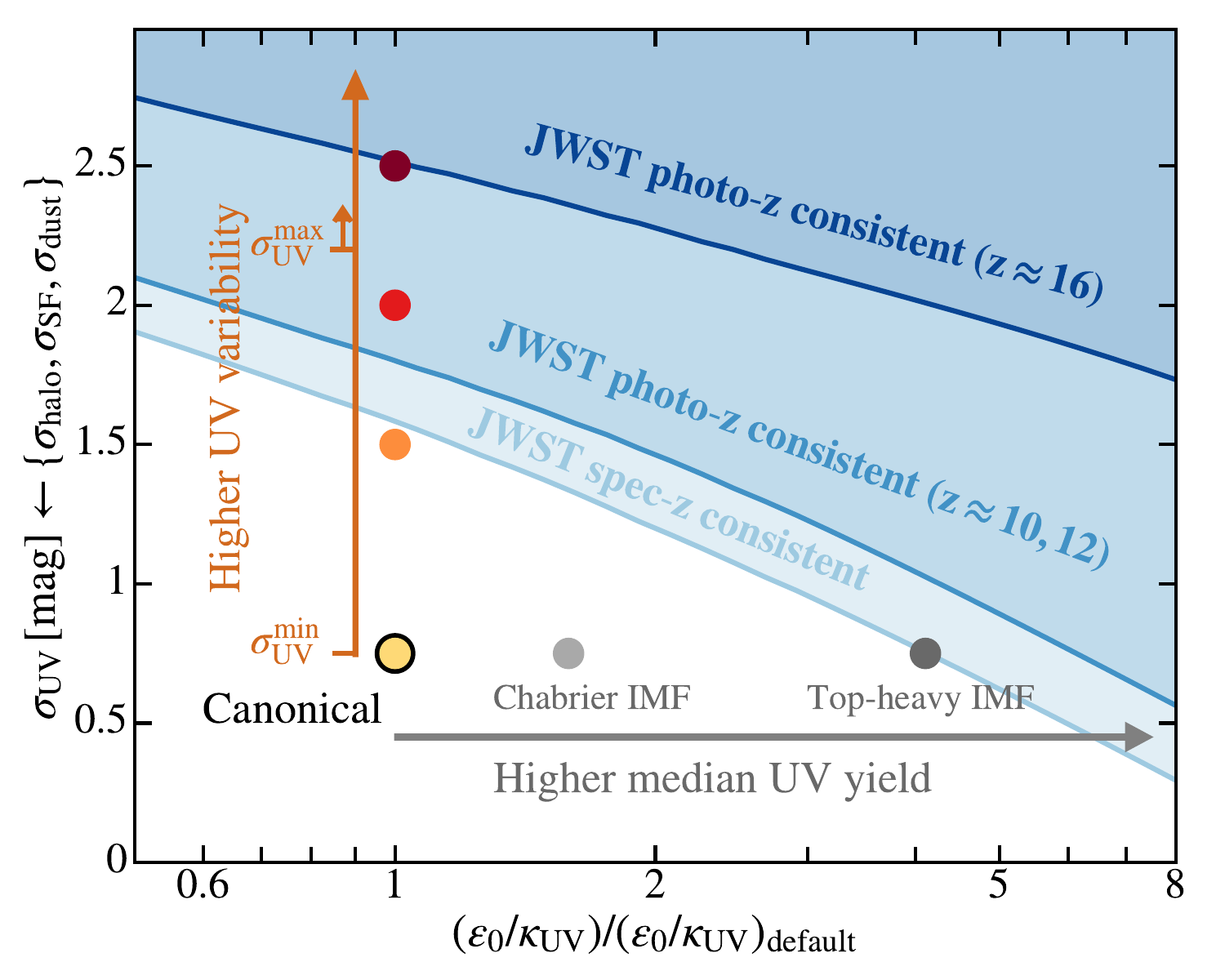}
    \caption{Parameter space of UV variability, $\sigma_{\rm UV}$, versus median UV radiation yield, represented by $\varepsilon_{0}/\kappa_{\rm UV}$. The shaded regions show the regions of parameter space consistent with the JWST results mapped by our empirical model. This parametrization highlights two distinct ways of easing the current tension between theoretical models and observations: enhancing the median UV radiation yield or enhancing the UV variability. The minimum and maximum $\sigma_{\rm UV}$ estimated in Section~\ref{sec:method:sigmaUV} are shown. Reasonable values of $\sigma_{\rm UV}$ within the constraints can explain the most stringent JWST results. In the horizontal direction, we show the enhancement of median UV radiation yield for e.g. two alternative IMFs.}
    \label{fig:parameters}
\end{figure}

To illustrate the implication of UV variability in reconciling JWST results with theoretical models, in Figure~\ref{fig:parameters}, we examine the parameter space of UV variability and the median UV radiation yield. For results based on JWST spectroscopy, we consider the model to be consistent with observations when $\log_{10}{\Phi}(M_{\rm UV}=-20.5)>-5$ and $-5.3$ at $z=$10 and 12, respectively. For photometric constraints, we consider the model to be acceptable when $\log_{10}{\Phi}(M_{\rm UV}=-20.5)>-4.7$, $-5.0$ and $-5.2$ at $z=$10, 12 and 16, respectively. We scan the parameter space by modifying UV variability and the normalization $\varepsilon_{0}$ of the star-formation efficiency in our model and identify the regime where theoretically predicted UV bright galaxy abundance exceeds the observed values. 

As illustrated in the figure, there are two ways to reconcile the model with the JWST results. One option is to enhance the \textit{median} UV radiation yield, either by boosting the star-formation efficiency or enhancing the UV radiation efficiency. For example, the $\kappa_{\rm UV}$ can drop significantly if assuming different IMFs. For a \citet{Chabrier2003} IMF, $\kappa_{\rm UV}$ can drop by roughly $37\%$~\citep{Madau2014}. For an extremely top-heavy IMF that may be appropriate for, e.g., metal-free Population {\small III} stars, $\kappa_{\rm UV}$ can drop by $76\%$, to $0.28\times 10^{-28}$~\citep{Inayoshi2022}. These scenarios are indicated by the gray dots in the figure. However, this approach can not reconcile the most stringent JWST photometric constraints at $z\approx 16$. An alternative approach is to enhance the variability in the observed UV luminosity, as highlighted by the colored dots and explored in more detail in earlier figures. In this paper, we have focused on the latter option and assumed a log-normal distribution of observed UV luminosity. However, in practice, a similar phenomenon can be driven by e.g. incorporating a fraction of starbursts with high star-formation efficiencies.

\section{Discussion and Conclusions}
\label{sec:conclusions}

In this paper, we study the impact of UV variability on the rest-frame UVLF of galaxies at $z\geq 9$ constrained by recent JWST observations. We introduce an empirical model that links host dark matter halo mass to the median galaxy UV luminosity and make predictions for the UVLF at high redshift. This model is designed to minimize dependence or assumptions on specific galaxy formation recipes and represent our understanding of galaxy formation prior to the JWST era. Based on this \textit{median} galaxy-halo connection, we investigate the extent of UV variability required to explain the substantial presence of UV-bright galaxies observed by JWST at $z \geq 9$. This UV variability encompasses the random fluctuations in halo assembly, star formation, and dust attenuation processes.

Even assuming \textit{all} the photometrically-selected candidates are real, we find that JWST observations at $z\approx 10$, $12$, $16$ can be reconciled with a standard galaxy formation model calibrated at low redshift with $\sigma_{\rm UV}\approx 1.5$, $2.0$, $2.5\mmag$. Our results indicate a transition at $z=10$. Below this redshift, $\sigma_{\rm UV} \approx 0.75 - 1.5$ mag is favored to match the UVLF and the cumulative UV luminosity density of the Universe. At higher redshifts, the required value of $\sigma_{\rm UV}$ is larger and grows with increasing redshift in order to reproduce the bright end of the UVLF. This transition implies a sharp change in the underlying mechanism that is responsible for the observed UV variability. UV emission is sensitive to the SFR over a time scale of $\sim 10-100\,\Myr$ \citep[e.g.][]{Murphy2011,Flores2021}, close to the dynamical timescale of a dark matter halo in virial equilibrium --- which sets the time scale of baryon cycles in high-redshift galaxies~ --- at $z\approx 10$~\citep[e.g.][]{Angles2017,Tacchella2020}. In addition, the characteristic redshift could correspond to the epoch when the cooling and free-fall time in dense gas disks becomes shorter than the time for low-metallicity massive stars to develop winds and supernovae~\citep[$\sim \Myr$;][]{FG2018,Dekel2023}. This scenario is explicitly studied in \citet{Dekel2023}, who found that feedback-free starburst with high star-formation efficiencies can occur at $z\gtrsim 10$.

The implied UV variability is consistent with the expected values from halo assembly, burstiness of star formation in high-redshift galaxies, and dust attenuation variations. In addition to using UV as the primary tracer, emission line measurements (e.g. H$\gamma$ and H$\delta$ using JWST NIRSpec, H$\alpha$ using JWST MIRI) for $z\approx 10$ galaxies will be useful in measuring the burstiness of star formation from, e.g., the ratio of H$\alpha$ versus UV luminosity~\citep[e.g.][]{Broussard2019,Caplar2019,Emami2019,Faisst2019,Iyer2022} and isolate the physical origin of the burstiness. These emission line tracers are sensitive to SFR as measured on very short time scales and are therefore useful for studying processes such as the feedback-free starbursts highlighted above and the typical lifecycle of giant molecular clouds~\citep[$\lesssim 10\Myr$; e.g.][]{Leitherer1999, Tan2000, Tasker2011}. High-resolution hydrodynamical simulations with \textbf{predictive power} below the interstellar medium (ISM) scale will also shed light on the physical origin of UV variability and its implication for resolving the UVLF tension~\citep{Pallottini2023,Sun2023}.

In summary, current theoretical frameworks such as empirical models, semi-analytical models, and large-volume numerical simulations might substantially underestimate the variability in UV luminosity of individual galaxies arising from various baryonic physics processes at or below the interstellar medium scale~\citep[e.g.][]{Iyer2020,Tacchella2020} in the extremely high-density environment at high redshift. These models underestimate the observed UVLF at high redshift unless they adopt an increase in the median UV radiation yield. However, by incorporating a physically motivated higher UV variability, the need for adjustments to a standard galaxy formation model --- such as introducing a top-heavy stellar initial mass function, a drastically different star-formation law, or considering significant contamination from non-stellar sources --- can be substantially reduced. As a result, the bright galaxy populations unveiled by JWST at $z\gtrsim 10$ are consistent with the $\Lambda$CDM cosmological model paired with a standard galaxy formation model, assuming a reasonable variability in UV luminosity. 

\section*{Acknowledgements}
We thank Josh Borrow, Rohan Naidu, Anna-Christina Eilers, Claude-Andr\'{e} Faucher-Gigu\`{e}re, Guochao Sun, and Christopher Hayward for suggestions that improved the manuscript. MV acknowledges support through NASA ATP 19-ATP19-0019, 19-ATP19-0020, 19-ATP19-0167, and NSF grants AST-1814053, AST-1814259, AST-1909831, AST-2007355 and AST-2107724. MBK acknowledges support from NSF CAREER award AST-1752913, NSF grants AST-1910346 and AST-2108962, NASA grant 80NSSC22K0827, and HST-AR-15809, HST-GO-15658, HST-GO-15901, HST-GO-15902, HST-AR-16159, HST-GO-16226, HST-GO-16686, HST-AR-17028, and HST-AR-17043 from the Space Telescope Science Institute, which is operated by AURA, Inc., under NASA contract NAS5-26555. 

\section*{Data Availability}
The data underlying this article can be shared on reasonable request to the corresponding author. The source code and the observational data compiled for this project are publically available at \href{https://github.com/XuejianShen/highz-empirical-variability}{the online repository}.




\begin{thebibliography}{}
\makeatletter
\relax
\def\mn@urlcharsother{\let\do\@makeother \do\$\do\&\do\#\do\^\do\_\do\%\do\~}
\def\mn@doi{\begingroup\mn@urlcharsother \@ifnextchar [ {\mn@doi@}
  {\mn@doi@[]}}
\def\mn@doi@[#1]#2{\def\@tempa{#1}\ifx\@tempa\@empty \href
  {http://dx.doi.org/#2} {doi:#2}\else \href {http://dx.doi.org/#2} {#1}\fi
  \endgroup}
\def\mn@eprint#1#2{\mn@eprint@#1:#2::\@nil}
\def\mn@eprint@arXiv#1{\href {http://arxiv.org/abs/#1} {{\tt arXiv:#1}}}
\def\mn@eprint@dblp#1{\href {http://dblp.uni-trier.de/rec/bibtex/#1.xml}
  {dblp:#1}}
\def\mn@eprint@#1:#2:#3:#4\@nil{\def\@tempa {#1}\def\@tempb {#2}\def\@tempc
  {#3}\ifx \@tempc \@empty \let \@tempc \@tempb \let \@tempb \@tempa \fi \ifx
  \@tempb \@empty \def\@tempb {arXiv}\fi \@ifundefined
  {mn@eprint@\@tempb}{\@tempb:\@tempc}{\expandafter \expandafter \csname
  mn@eprint@\@tempb\endcsname \expandafter{\@tempc}}}

\bibitem[\protect\citeauthoryear{{Adams} et~al.,}{{Adams}
  et~al.}{2023a}]{Adams2023b}
{Adams} N.~J.,  et~al., 2023a, \mn@doi [arXiv e-prints]
  {10.48550/arXiv.2304.13721}, \href
  {https://ui.adsabs.harvard.edu/abs/2023arXiv230413721A} {p. arXiv:2304.13721}

\bibitem[\protect\citeauthoryear{{Adams} et~al.,}{{Adams}
  et~al.}{2023b}]{Adams2023}
{Adams} N.~J.,  et~al., 2023b, \mn@doi [\mnras] {10.1093/mnras/stac3347}, \href
  {https://ui.adsabs.harvard.edu/abs/2023MNRAS.518.4755A} {518, 4755}

\bibitem[\protect\citeauthoryear{{Angl{\'e}s-Alc{\'a}zar},
  {Faucher-Gigu{\`e}re}, {Kere{\v{s}}}, {Hopkins}, {Quataert}  \&
  {Murray}}{{Angl{\'e}s-Alc{\'a}zar} et~al.}{2017}]{Angles2017}
{Angl{\'e}s-Alc{\'a}zar} D.,  {Faucher-Gigu{\`e}re} C.-A.,  {Kere{\v{s}}} D.,
  {Hopkins} P.~F.,  {Quataert} E.,   {Murray} N.,  2017, \mn@doi [\mnras]
  {10.1093/mnras/stx1517}, \href
  {https://ui.adsabs.harvard.edu/abs/2017MNRAS.470.4698A} {470, 4698}

\bibitem[\protect\citeauthoryear{{Arrabal Haro} et~al.,}{{Arrabal Haro}
  et~al.}{2023a}]{AH2023b}
{Arrabal Haro} P.,  et~al., 2023a, \mn@doi [arXiv e-prints]
  {10.48550/arXiv.2303.15431}, \href
  {https://ui.adsabs.harvard.edu/abs/2023arXiv230315431A} {p. arXiv:2303.15431}

\bibitem[\protect\citeauthoryear{{Arrabal Haro} et~al.,}{{Arrabal Haro}
  et~al.}{2023b}]{AH2023a}
{Arrabal Haro} P.,  et~al., 2023b, \mn@doi [arXiv e-prints]
  {10.48550/arXiv.2304.05378}, \href
  {https://ui.adsabs.harvard.edu/abs/2023arXiv230405378A} {p. arXiv:2304.05378}

\bibitem[\protect\citeauthoryear{{Atek} et~al.,}{{Atek}
  et~al.}{2023}]{Atek2023}
{Atek} H.,  et~al., 2023, \mn@doi [\mnras] {10.1093/mnras/stac3144}, \href
  {https://ui.adsabs.harvard.edu/abs/2023MNRAS.519.1201A} {519, 1201}

\bibitem[\protect\citeauthoryear{{Behroozi}, {Wechsler}, {Hearin}  \&
  {Conroy}}{{Behroozi} et~al.}{2019}]{Behroozi2019}
{Behroozi} P.,  {Wechsler} R.~H.,  {Hearin} A.~P.,   {Conroy} C.,  2019,
  \mn@doi [\mnras] {10.1093/mnras/stz1182}, \href
  {https://ui.adsabs.harvard.edu/abs/2019MNRAS.488.3143B} {488, 3143}

\bibitem[\protect\citeauthoryear{{Behroozi} et~al.,}{{Behroozi}
  et~al.}{2020}]{Behroozi2020}
{Behroozi} P.,  et~al., 2020, \mn@doi [\mnras] {10.1093/mnras/staa3164}, \href
  {https://ui.adsabs.harvard.edu/abs/2020MNRAS.499.5702B} {499, 5702}

\bibitem[\protect\citeauthoryear{{Bond}, {Cole}, {Efstathiou}  \&
  {Kaiser}}{{Bond} et~al.}{1991}]{Bond1991}
{Bond} J.~R.,  {Cole} S.,  {Efstathiou} G.,   {Kaiser} N.,  1991, \mn@doi
  [\apj] {10.1086/170520}, \href
  {https://ui.adsabs.harvard.edu/abs/1991ApJ...379..440B} {379, 440}

\bibitem[\protect\citeauthoryear{Bournaud, Elmegreen  \& Elmegreen}{Bournaud
  et~al.}{2007}]{Bournaud2007}
Bournaud F.,  Elmegreen B.~G.,   Elmegreen D.~M.,  2007, \mn@doi [The
  Astrophysical Journal] {10.1086/522077}, 670, 237

\bibitem[\protect\citeauthoryear{{Bouwens} et~al.,}{{Bouwens}
  et~al.}{2014}]{Bouwens2014}
{Bouwens} R.~J.,  et~al., 2014, \mn@doi [\apj] {10.1088/0004-637X/793/2/115},
  \href {https://ui.adsabs.harvard.edu/abs/2014ApJ...793..115B} {793, 115}

\bibitem[\protect\citeauthoryear{{Bouwens}, {Illingworth}, {Oesch}, {Caruana},
  {Holwerda}, {Smit}  \& {Wilkins}}{{Bouwens} et~al.}{2015}]{Bouwens2015}
{Bouwens} R.~J.,  {Illingworth} G.~D.,  {Oesch} P.~A.,  {Caruana} J.,
  {Holwerda} B.,  {Smit} R.,   {Wilkins} S.,  2015, \mn@doi [\apj]
  {10.1088/0004-637X/811/2/140}, \href
  {https://ui.adsabs.harvard.edu/abs/2015ApJ...811..140B} {811, 140}

\bibitem[\protect\citeauthoryear{{Bouwens} et~al.,}{{Bouwens}
  et~al.}{2020}]{Bouwens2020}
{Bouwens} R.,  et~al., 2020, \mn@doi [\apj] {10.3847/1538-4357/abb830}, \href
  {https://ui.adsabs.harvard.edu/abs/2020ApJ...902..112B} {902, 112}

\bibitem[\protect\citeauthoryear{{Bouwens} et~al.,}{{Bouwens}
  et~al.}{2021}]{Bouwens2021}
{Bouwens} R.~J.,  et~al., 2021, \mn@doi [\aj] {10.3847/1538-3881/abf83e}, \href
  {https://ui.adsabs.harvard.edu/abs/2021AJ....162...47B} {162, 47}

\bibitem[\protect\citeauthoryear{{Bouwens} et~al.,}{{Bouwens}
  et~al.}{2023a}]{Bouwens2023a}
{Bouwens} R.~J.,  et~al., 2023a, \mn@doi [\mnras] {10.1093/mnras/stad1145},
  \href {https://ui.adsabs.harvard.edu/abs/2023MNRAS.tmp.1093B} {}

\bibitem[\protect\citeauthoryear{{Bouwens}, {Illingworth}, {Oesch}, {Stefanon},
  {Naidu}, {van Leeuwen}  \& {Magee}}{{Bouwens} et~al.}{2023b}]{Bouwens2023b}
{Bouwens} R.,  {Illingworth} G.,  {Oesch} P.,  {Stefanon} M.,  {Naidu} R.,
  {van Leeuwen} I.,   {Magee} D.,  2023b, \mn@doi [\mnras]
  {10.1093/mnras/stad1014}, \href
  {https://ui.adsabs.harvard.edu/abs/2023MNRAS.tmp.1019B} {}

\bibitem[\protect\citeauthoryear{{Bowler}, {Jarvis}, {Dunlop}, {McLure},
  {McLeod}, {Adams}, {Milvang-Jensen}  \& {McCracken}}{{Bowler}
  et~al.}{2020}]{Bowler2020}
{Bowler} R.~A.~A.,  {Jarvis} M.~J.,  {Dunlop} J.~S.,  {McLure} R.~J.,  {McLeod}
  D.~J.,  {Adams} N.~J.,  {Milvang-Jensen} B.,   {McCracken} H.~J.,  2020,
  \mn@doi [\mnras] {10.1093/mnras/staa313}, \href
  {https://ui.adsabs.harvard.edu/abs/2020MNRAS.493.2059B} {493, 2059}

\bibitem[\protect\citeauthoryear{{Boylan-Kolchin}}{{Boylan-Kolchin}}{2023}]{BK2023}
{Boylan-Kolchin} M.,  2023, \mn@doi [\natastro] {10.1038/s41550-023-01937-7},
  \href {https://ui.adsabs.harvard.edu/abs/2023NatAs.tmp...77B} {pp
  10.1038/s41550--023--01937--7}

\bibitem[\protect\citeauthoryear{{Boylan-Kolchin}, {Springel}, {White},
  {Jenkins}  \& {Lemson}}{{Boylan-Kolchin} et~al.}{2009}]{BK2009}
{Boylan-Kolchin} M.,  {Springel} V.,  {White} S. D.~M.,  {Jenkins} A.,
  {Lemson} G.,  2009, \mn@doi [\mnras] {10.1111/j.1365-2966.2009.15191.x},
  \href {https://ui.adsabs.harvard.edu/abs/2009MNRAS.398.1150B} {398, 1150}

\bibitem[\protect\citeauthoryear{{Broussard} et~al.,}{{Broussard}
  et~al.}{2019}]{Broussard2019}
{Broussard} A.,  et~al., 2019, \mn@doi [\apj] {10.3847/1538-4357/ab04ad}, \href
  {https://ui.adsabs.harvard.edu/abs/2019ApJ...873...74B} {873, 74}

\bibitem[\protect\citeauthoryear{{Bryan} \& {Norman}}{{Bryan} \&
  {Norman}}{1998}]{Bryan1998}
{Bryan} G.~L.,  {Norman} M.~L.,  1998, \mn@doi [\apj] {10.1086/305262}, \href
  {https://ui.adsabs.harvard.edu/abs/1998ApJ...495...80B} {495, 80}

\bibitem[\protect\citeauthoryear{{Bunker} et~al.,}{{Bunker}
  et~al.}{2023}]{Bunker2023}
{Bunker} A.~J.,  et~al., 2023, \mn@doi [arXiv e-prints]
  {10.48550/arXiv.2302.07256}, \href
  {https://ui.adsabs.harvard.edu/abs/2023arXiv230207256B} {p. arXiv:2302.07256}

\bibitem[\protect\citeauthoryear{{Caplar} \& {Tacchella}}{{Caplar} \&
  {Tacchella}}{2019}]{Caplar2019}
{Caplar} N.,  {Tacchella} S.,  2019, \mn@doi [\mnras] {10.1093/mnras/stz1449},
  \href {https://ui.adsabs.harvard.edu/abs/2019MNRAS.487.3845C} {487, 3845}

\bibitem[\protect\citeauthoryear{{Carniani} et~al.,}{{Carniani}
  et~al.}{2018}]{Carniani2018}
{Carniani} S.,  et~al., 2018, \mn@doi [\mnras] {10.1093/mnras/sty1088}, \href
  {https://ui.adsabs.harvard.edu/abs/2018MNRAS.478.1170C} {478, 1170}

\bibitem[\protect\citeauthoryear{{Castellano} et~al.,}{{Castellano}
  et~al.}{2022}]{Castellano2022}
{Castellano} M.,  et~al., 2022, \mn@doi [\apjl] {10.3847/2041-8213/ac94d0},
  \href {https://ui.adsabs.harvard.edu/abs/2022ApJ...938L..15C} {938, L15}

\bibitem[\protect\citeauthoryear{{Ceverino}, {Dekel}  \& {Bournaud}}{{Ceverino}
  et~al.}{2010}]{Ceverino2010}
{Ceverino} D.,  {Dekel} A.,   {Bournaud} F.,  2010, \mn@doi [\mnras]
  {10.1111/j.1365-2966.2010.16433.x}, \href
  {https://ui.adsabs.harvard.edu/abs/2010MNRAS.404.2151C} {404, 2151}

\bibitem[\protect\citeauthoryear{{Chabrier}}{{Chabrier}}{2003}]{Chabrier2003}
{Chabrier} G.,  2003, \mn@doi [\pasp] {10.1086/376392}, \href
  {https://ui.adsabs.harvard.edu/abs/2003PASP..115..763C} {115, 763}

\bibitem[\protect\citeauthoryear{{Chen}, {Mo}  \& {Wang}}{{Chen}
  et~al.}{2023}]{Chen2023}
{Chen} Y.,  {Mo} H.~J.,   {Wang} K.,  2023, \mn@doi [arXiv e-prints]
  {10.48550/arXiv.2304.13890}, \href
  {https://ui.adsabs.harvard.edu/abs/2023arXiv230413890C} {p. arXiv:2304.13890}

\bibitem[\protect\citeauthoryear{{Cochrane} et~al.,}{{Cochrane}
  et~al.}{2019}]{Cochrane2019}
{Cochrane} R.~K.,  et~al., 2019, \mn@doi [\mnras] {10.1093/mnras/stz1736},
  \href {https://ui.adsabs.harvard.edu/abs/2019MNRAS.488.1779C} {488, 1779}

\bibitem[\protect\citeauthoryear{{Coe} et~al.,}{{Coe} et~al.}{2013}]{Coe2013}
{Coe} D.,  et~al., 2013, \mn@doi [\apj] {10.1088/0004-637X/762/1/32}, \href
  {https://ui.adsabs.harvard.edu/abs/2013ApJ...762...32C} {762, 32}

\bibitem[\protect\citeauthoryear{{Cullen} et~al.,}{{Cullen}
  et~al.}{2023}]{Cullen2023}
{Cullen} F.,  et~al., 2023, \mn@doi [\mnras] {10.1093/mnras/stad073}, \href
  {https://ui.adsabs.harvard.edu/abs/2023MNRAS.520...14C} {520, 14}

\bibitem[\protect\citeauthoryear{{Curtis-Lake} et~al.,}{{Curtis-Lake}
  et~al.}{2023}]{Curtis2023}
{Curtis-Lake} E.,  et~al., 2023, \mn@doi [Nature Astronomy]
  {10.1038/s41550-023-01918-w}, \href
  {https://ui.adsabs.harvard.edu/abs/2023NatAs.tmp...66C} {}

\bibitem[\protect\citeauthoryear{{Dav{\'e}}, {Angl{\'e}s-Alc{\'a}zar},
  {Narayanan}, {Li}, {Rafieferantsoa}  \& {Appleby}}{{Dav{\'e}}
  et~al.}{2019}]{Simba}
{Dav{\'e}} R.,  {Angl{\'e}s-Alc{\'a}zar} D.,  {Narayanan} D.,  {Li} Q.,
  {Rafieferantsoa} M.~H.,   {Appleby} S.,  2019, \mn@doi [\mnras]
  {10.1093/mnras/stz937}, \href
  {https://ui.adsabs.harvard.edu/abs/2019MNRAS.486.2827D} {486, 2827}

\bibitem[\protect\citeauthoryear{{Dayal}, {Ferrara}, {Dunlop}  \&
  {Pacucci}}{{Dayal} et~al.}{2014}]{Dayal2014}
{Dayal} P.,  {Ferrara} A.,  {Dunlop} J.~S.,   {Pacucci} F.,  2014, \mn@doi
  [\mnras] {10.1093/mnras/stu1848}, \href
  {https://ui.adsabs.harvard.edu/abs/2014MNRAS.445.2545D} {445, 2545}

\bibitem[\protect\citeauthoryear{{Dayal}, {Rossi}, {Shiralilou}, {Piana},
  {Choudhury}  \& {Volonteri}}{{Dayal} et~al.}{2019}]{Dayal2019}
{Dayal} P.,  {Rossi} E.~M.,  {Shiralilou} B.,  {Piana} O.,  {Choudhury} T.~R.,
   {Volonteri} M.,  2019, \mn@doi [\mnras] {10.1093/mnras/stz897}, \href
  {https://ui.adsabs.harvard.edu/abs/2019MNRAS.486.2336D} {486, 2336}

\bibitem[\protect\citeauthoryear{{Dekel}, {Sari}  \& {Ceverino}}{{Dekel}
  et~al.}{2009}]{Dekel2009}
{Dekel} A.,  {Sari} R.,   {Ceverino} D.,  2009, \mn@doi [\apj]
  {10.1088/0004-637X/703/1/785}, \href
  {https://ui.adsabs.harvard.edu/abs/2009ApJ...703..785D} {703, 785}

\bibitem[\protect\citeauthoryear{{Dekel}, {Sarkar}, {Birnboim}, {Mandelker}  \&
  {Li}}{{Dekel} et~al.}{2023}]{Dekel2023}
{Dekel} A.,  {Sarkar} K.~C.,  {Birnboim} Y.,  {Mandelker} N.,   {Li} Z.,  2023,
  \mn@doi [\mnras] {10.1093/mnras/stad1557}, \href
  {https://ui.adsabs.harvard.edu/abs/2023MNRAS.523.3201D} {523, 3201}

\bibitem[\protect\citeauthoryear{{Dong}, {Zhao}, {Han}, {Li}, {Jing}  \&
  {Yang}}{{Dong} et~al.}{2022}]{Dong2022}
{Dong} F.,  {Zhao} D.,  {Han} J.,  {Li} Z.,  {Jing} Y.,   {Yang} X.,  2022,
  \mn@doi [\apj] {10.3847/1538-4357/ac5aaa}, \href
  {https://ui.adsabs.harvard.edu/abs/2022ApJ...929..120D} {929, 120}

\bibitem[\protect\citeauthoryear{{Donnan} et~al.,}{{Donnan}
  et~al.}{2023}]{Donnan2023}
{Donnan} C.~T.,  et~al., 2023, \mn@doi [\mnras] {10.1093/mnras/stac3472}, \href
  {https://ui.adsabs.harvard.edu/abs/2023MNRAS.518.6011D} {518, 6011}

\bibitem[\protect\citeauthoryear{{El-Badry}, {Wetzel}, {Geha}, {Hopkins},
  {Kere{\v{s}}}, {Chan}  \& {Faucher-Gigu{\`e}re}}{{El-Badry}
  et~al.}{2016}]{ElBadry2016}
{El-Badry} K.,  {Wetzel} A.,  {Geha} M.,  {Hopkins} P.~F.,  {Kere{\v{s}}} D.,
  {Chan} T.~K.,   {Faucher-Gigu{\`e}re} C.-A.,  2016, \mn@doi [\apj]
  {10.3847/0004-637X/820/2/131}, \href
  {https://ui.adsabs.harvard.edu/abs/2016ApJ...820..131E} {820, 131}

\bibitem[\protect\citeauthoryear{{Ellis} et~al.,}{{Ellis}
  et~al.}{2013}]{Ellis2013}
{Ellis} R.~S.,  et~al., 2013, \mn@doi [\apjl] {10.1088/2041-8205/763/1/L7},
  \href {https://ui.adsabs.harvard.edu/abs/2013ApJ...763L...7E} {763, L7}

\bibitem[\protect\citeauthoryear{{Elmegreen}, {Elmegreen}, {Fernandez}  \&
  {Lemonias}}{{Elmegreen} et~al.}{2009}]{Elmegreen2009}
{Elmegreen} B.~G.,  {Elmegreen} D.~M.,  {Fernandez} M.~X.,   {Lemonias} J.~J.,
  2009, \mn@doi [\apj] {10.1088/0004-637X/692/1/12}, \href
  {http://adsabs.harvard.edu/abs/2009ApJ...692...12E} {692, 12}

\bibitem[\protect\citeauthoryear{{Emami}, {Siana}, {Weisz}, {Johnson}, {Ma}  \&
  {El-Badry}}{{Emami} et~al.}{2019}]{Emami2019}
{Emami} N.,  {Siana} B.,  {Weisz} D.~R.,  {Johnson} B.~D.,  {Ma} X.,
  {El-Badry} K.,  2019, \mn@doi [\apj] {10.3847/1538-4357/ab211a}, \href
  {https://ui.adsabs.harvard.edu/abs/2019ApJ...881...71E} {881, 71}

\bibitem[\protect\citeauthoryear{{Endsley}, {Stark}, {Whitler}, {Topping},
  {Chen}, {Plat}, {Chisholm}  \& {Charlot}}{{Endsley}
  et~al.}{2023}]{Endsley2022}
{Endsley} R.,  {Stark} D.~P.,  {Whitler} L.,  {Topping} M.~W.,  {Chen} Z.,
  {Plat} A.,  {Chisholm} J.,   {Charlot} S.,  2023, \mn@doi [\mnras]
  {10.1093/mnras/stad1919}, \href
  {https://ui.adsabs.harvard.edu/abs/2023MNRAS.524.2312E} {524, 2312}

\bibitem[\protect\citeauthoryear{{Faisst}, {Capak}, {Emami}, {Tacchella}  \&
  {Larson}}{{Faisst} et~al.}{2019}]{Faisst2019}
{Faisst} A.~L.,  {Capak} P.~L.,  {Emami} N.,  {Tacchella} S.,   {Larson} K.~L.,
   2019, \mn@doi [\apj] {10.3847/1538-4357/ab425b}, \href
  {https://ui.adsabs.harvard.edu/abs/2019ApJ...884..133F} {884, 133}

\bibitem[\protect\citeauthoryear{{Fakhouri}, {Ma}  \&
  {Boylan-Kolchin}}{{Fakhouri} et~al.}{2010}]{Fakhouri2010}
{Fakhouri} O.,  {Ma} C.-P.,   {Boylan-Kolchin} M.,  2010, \mn@doi [\mnras]
  {10.1111/j.1365-2966.2010.16859.x}, \href
  {https://ui.adsabs.harvard.edu/abs/2010MNRAS.406.2267F} {406, 2267}

\bibitem[\protect\citeauthoryear{{Faucher-Gigu{\`e}re}}{{Faucher-Gigu{\`e}re}}{2018}]{FG2018}
{Faucher-Gigu{\`e}re} C.-A.,  2018, \mn@doi [\mnras] {10.1093/mnras/stx2595},
  \href {https://ui.adsabs.harvard.edu/abs/2018MNRAS.473.3717F} {473, 3717}

\bibitem[\protect\citeauthoryear{{Ferrara} et~al.,}{{Ferrara}
  et~al.}{2022}]{Ferrara2022b}
{Ferrara} A.,  et~al., 2022, \mn@doi [\mnras] {10.1093/mnras/stac460}, \href
  {https://ui.adsabs.harvard.edu/abs/2022MNRAS.512...58F} {512, 58}

\bibitem[\protect\citeauthoryear{{Ferrara}, {Pallottini}  \& {Dayal}}{{Ferrara}
  et~al.}{2023}]{Ferrara2022}
{Ferrara} A.,  {Pallottini} A.,   {Dayal} P.,  2023, \mn@doi [\mnras]
  {10.1093/mnras/stad1095}, \href
  {https://ui.adsabs.harvard.edu/abs/2023MNRAS.522.3986F} {522, 3986}

\bibitem[\protect\citeauthoryear{{Finkelstein} et~al.,}{{Finkelstein}
  et~al.}{2022}]{Finkelstein2022}
{Finkelstein} S.~L.,  et~al., 2022, \mn@doi [\apjl] {10.3847/2041-8213/ac966e},
  \href {https://ui.adsabs.harvard.edu/abs/2022ApJ...940L..55F} {940, L55}

\bibitem[\protect\citeauthoryear{{Finkelstein} et~al.,}{{Finkelstein}
  et~al.}{2023}]{Finkelstein2023}
{Finkelstein} S.~L.,  et~al., 2023, \mn@doi [\apjl] {10.3847/2041-8213/acade4},
  \href {https://ui.adsabs.harvard.edu/abs/2023ApJ...946L..13F} {946, L13}

\bibitem[\protect\citeauthoryear{{Fiore}, {Ferrara}, {Bischetti}, {Feruglio}
  \& {Travascio}}{{Fiore} et~al.}{2023}]{Flore2023}
{Fiore} F.,  {Ferrara} A.,  {Bischetti} M.,  {Feruglio} C.,   {Travascio} A.,
  2023, \mn@doi [\apjl] {10.3847/2041-8213/acb5f2}, \href
  {https://ui.adsabs.harvard.edu/abs/2023ApJ...943L..27F} {943, L27}

\bibitem[\protect\citeauthoryear{{Flores Vel{\'a}zquez} et~al.,}{{Flores
  Vel{\'a}zquez} et~al.}{2021}]{Flores2021}
{Flores Vel{\'a}zquez} J.~A.,  et~al., 2021, \mn@doi [\mnras]
  {10.1093/mnras/staa3893}, \href
  {https://ui.adsabs.harvard.edu/abs/2021MNRAS.501.4812F} {501, 4812}

\bibitem[\protect\citeauthoryear{{F{\"o}rster Schreiber} et~al.}{{F{\"o}rster
  Schreiber} et~al.}{2011}]{Forster2011}
{F{\"o}rster Schreiber} N.~M.,  et~al., 2011, \mn@doi [\apj]
  {10.1088/0004-637X/731/1/65}, \href
  {http://adsabs.harvard.edu/abs/2011ApJ...731...65F} {731, 65}

\bibitem[\protect\citeauthoryear{{Fujimoto} et~al.,}{{Fujimoto}
  et~al.}{2022}]{Fujimoto2022}
{Fujimoto} S.,  et~al., 2022, \mn@doi [arXiv e-prints]
  {10.48550/arXiv.2211.03896}, \href
  {https://ui.adsabs.harvard.edu/abs/2022arXiv221103896F} {p. arXiv:2211.03896}

\bibitem[\protect\citeauthoryear{{Harikane} et~al.,}{{Harikane}
  et~al.}{2018}]{Harikane2018}
{Harikane} Y.,  et~al., 2018, \mn@doi [\pasj] {10.1093/pasj/psx097}, \href
  {https://ui.adsabs.harvard.edu/abs/2018PASJ...70S..11H} {70, S11}

\bibitem[\protect\citeauthoryear{{Harikane} et~al.,}{{Harikane}
  et~al.}{2022}]{Harikane2022}
{Harikane} Y.,  et~al., 2022, \mn@doi [\apjs] {10.3847/1538-4365/ac3dfc}, \href
  {https://ui.adsabs.harvard.edu/abs/2022ApJS..259...20H} {259, 20}

\bibitem[\protect\citeauthoryear{{Harikane}, {Nakajima}, {Ouchi}, {Umeda},
  {Isobe}, {Ono}, {Xu}  \& {Zhang}}{{Harikane}
  et~al.}{2023a}]{Harikane2023-spec}
{Harikane} Y.,  {Nakajima} K.,  {Ouchi} M.,  {Umeda} H.,  {Isobe} Y.,  {Ono}
  Y.,  {Xu} Y.,   {Zhang} Y.,  2023a, \mn@doi [arXiv e-prints]
  {10.48550/arXiv.2304.06658}, \href
  {https://ui.adsabs.harvard.edu/abs/2023arXiv230406658H} {p. arXiv:2304.06658}

\bibitem[\protect\citeauthoryear{{Harikane} et~al.,}{{Harikane}
  et~al.}{2023b}]{Harikane2023}
{Harikane} Y.,  et~al., 2023b, \mn@doi [\apjs] {10.3847/1538-4365/acaaa9},
  \href {https://ui.adsabs.harvard.edu/abs/2023ApJS..265....5H} {265, 5}

\bibitem[\protect\citeauthoryear{{Haslbauer}, {Kroupa}, {Zonoozi}  \&
  {Haghi}}{{Haslbauer} et~al.}{2022}]{Haslbauer2022}
{Haslbauer} M.,  {Kroupa} P.,  {Zonoozi} A.~H.,   {Haghi} H.,  2022, \mn@doi
  [\apjl] {10.3847/2041-8213/ac9a50}, \href
  {https://ui.adsabs.harvard.edu/abs/2022ApJ...939L..31H} {939, L31}

\bibitem[\protect\citeauthoryear{{Hopkins} et~al.,}{{Hopkins}
  et~al.}{2023}]{Hopkins2023}
{Hopkins} P.~F.,  et~al., 2023, \mn@doi [\mnras] {10.1093/mnras/stad1902},
  \href {https://ui.adsabs.harvard.edu/abs/2023MNRAS.tmp.1847H} {}

\bibitem[\protect\citeauthoryear{Howlett, Lewis, Hall  \& Challinor}{Howlett
  et~al.}{2012}]{CAMB2}
Howlett C.,  Lewis A.,  Hall A.,   Challinor A.,  2012, \mn@doi [\jcap]
  {10.1088/1475-7516/2012/04/027}, 1204, 027

\bibitem[\protect\citeauthoryear{{Inayoshi}, {Harikane}, {Inoue}, {Li}  \&
  {Ho}}{{Inayoshi} et~al.}{2022}]{Inayoshi2022}
{Inayoshi} K.,  {Harikane} Y.,  {Inoue} A.~K.,  {Li} W.,   {Ho} L.~C.,  2022,
  \mn@doi [\apjl] {10.3847/2041-8213/ac9310}, \href
  {https://ui.adsabs.harvard.edu/abs/2022ApJ...938L..10I} {938, L10}

\bibitem[\protect\citeauthoryear{{Iyer} et~al.,}{{Iyer}
  et~al.}{2020}]{Iyer2020}
{Iyer} K.~G.,  et~al., 2020, \mn@doi [\mnras] {10.1093/mnras/staa2150}, \href
  {https://ui.adsabs.harvard.edu/abs/2020MNRAS.498..430I} {498, 430}

\bibitem[\protect\citeauthoryear{{Iyer}, {Speagle}, {Caplar}, {Forbes},
  {Gawiser}, {Leja}  \& {Tacchella}}{{Iyer} et~al.}{2022}]{Iyer2022}
{Iyer} K.~G.,  {Speagle} J.~S.,  {Caplar} N.,  {Forbes} J.~C.,  {Gawiser} E.,
  {Leja} J.,   {Tacchella} S.,  2022, \mn@doi [arXiv e-prints]
  {10.48550/arXiv.2208.05938}, \href
  {https://ui.adsabs.harvard.edu/abs/2022arXiv220805938I} {p. arXiv:2208.05938}

\bibitem[\protect\citeauthoryear{{Kannan}, {Garaldi}, {Smith}, {Pakmor},
  {Springel}, {Vogelsberger}  \& {Hernquist}}{{Kannan}
  et~al.}{2022}]{Kannan2022-thesan}
{Kannan} R.,  {Garaldi} E.,  {Smith} A.,  {Pakmor} R.,  {Springel} V.,
  {Vogelsberger} M.,   {Hernquist} L.,  2022, \mn@doi [\mnras]
  {10.1093/mnras/stab3710}, \href
  {https://ui.adsabs.harvard.edu/abs/2022MNRAS.511.4005K} {511, 4005}

\bibitem[\protect\citeauthoryear{{Kannan} et~al.,}{{Kannan}
  et~al.}{2023}]{Kannan2022}
{Kannan} R.,  et~al., 2023, \mn@doi [\mnras] {10.1093/mnras/stac3743}, \href
  {https://ui.adsabs.harvard.edu/abs/2023MNRAS.524.2594K} {524, 2594}

\bibitem[\protect\citeauthoryear{{Keller}, {Munshi}, {Trebitsch}  \&
  {Tremmel}}{{Keller} et~al.}{2023}]{Keller2023}
{Keller} B.~W.,  {Munshi} F.,  {Trebitsch} M.,   {Tremmel} M.,  2023, \mn@doi
  [\apjl] {10.3847/2041-8213/acb148}, \href
  {https://ui.adsabs.harvard.edu/abs/2023ApJ...943L..28K} {943, L28}

\bibitem[\protect\citeauthoryear{{Labb{\'e}} et~al.,}{{Labb{\'e}}
  et~al.}{2023}]{Labbe2023}
{Labb{\'e}} I.,  et~al., 2023, \mn@doi [\nat] {10.1038/s41586-023-05786-2},
  \href {https://ui.adsabs.harvard.edu/abs/2023Natur.616..266L} {616, 266}

\bibitem[\protect\citeauthoryear{{Larson} et~al.,}{{Larson}
  et~al.}{2022}]{Larson2022}
{Larson} R.~L.,  et~al., 2022, \mn@doi [arXiv e-prints]
  {10.48550/arXiv.2211.10035}, \href
  {https://ui.adsabs.harvard.edu/abs/2022arXiv221110035L} {p. arXiv:2211.10035}

\bibitem[\protect\citeauthoryear{{Leethochawalit} et~al.,}{{Leethochawalit}
  et~al.}{2023}]{Leetho2023}
{Leethochawalit} N.,  et~al., 2023, \mn@doi [\apjl] {10.3847/2041-8213/ac959b},
  \href {https://ui.adsabs.harvard.edu/abs/2023ApJ...942L..26L} {942, L26}

\bibitem[\protect\citeauthoryear{{Leitherer} et~al.,}{{Leitherer}
  et~al.}{1999}]{Leitherer1999}
{Leitherer} C.,  et~al., 1999, \mn@doi [\apjs] {10.1086/313233}, \href
  {https://ui.adsabs.harvard.edu/abs/1999ApJS..123....3L} {123, 3}

\bibitem[\protect\citeauthoryear{Lewis, Challinor  \& Lasenby}{Lewis
  et~al.}{2000}]{CAMB1}
Lewis A.,  Challinor A.,   Lasenby A.,  2000, \mn@doi [\apj] {10.1086/309179},
  538, 473

\bibitem[\protect\citeauthoryear{{Lovell}, {Harrison}, {Harikane}, {Tacchella}
  \& {Wilkins}}{{Lovell} et~al.}{2023}]{Lovell2023}
{Lovell} C.~C.,  {Harrison} I.,  {Harikane} Y.,  {Tacchella} S.,   {Wilkins}
  S.~M.,  2023, \mn@doi [\mnras] {10.1093/mnras/stac3224}, \href
  {https://ui.adsabs.harvard.edu/abs/2023MNRAS.518.2511L} {518, 2511}

\bibitem[\protect\citeauthoryear{{Madau} \& {Dickinson}}{{Madau} \&
  {Dickinson}}{2014}]{Madau2014}
{Madau} P.,  {Dickinson} M.,  2014, \mn@doi [\araa]
  {10.1146/annurev-astro-081811-125615}, \href
  {https://ui.adsabs.harvard.edu/abs/2014ARA&A..52..415M} {52, 415}

\bibitem[\protect\citeauthoryear{Mason, Trenti  \& Treu}{Mason
  et~al.}{2015}]{Mason2015}
Mason C.~A.,  Trenti M.,   Treu T.,  2015, \mn@doi [The Astrophysical Journal]
  {10.1088/0004-637X/813/1/21}, 813, 21

\bibitem[\protect\citeauthoryear{{Mason}, {Trenti}  \& {Treu}}{{Mason}
  et~al.}{2023}]{Mason2023}
{Mason} C.~A.,  {Trenti} M.,   {Treu} T.,  2023, \mn@doi [\mnras]
  {10.1093/mnras/stad035}, \href
  {https://ui.adsabs.harvard.edu/abs/2023MNRAS.521..497M} {521, 497}

\bibitem[\protect\citeauthoryear{{Mauerhofer} \& {Dayal}}{{Mauerhofer} \&
  {Dayal}}{2023}]{Mauerhofer2023}
{Mauerhofer} V.,  {Dayal} P.,  2023, \mn@doi [arXiv e-prints]
  {10.48550/arXiv.2305.01681}, \href
  {https://ui.adsabs.harvard.edu/abs/2023arXiv230501681M} {p. arXiv:2305.01681}

\bibitem[\protect\citeauthoryear{{McCaffrey}, {Hardin}, {Wise}  \&
  {Regan}}{{McCaffrey} et~al.}{2023}]{McCaffrey2023}
{McCaffrey} J.,  {Hardin} S.,  {Wise} J.,   {Regan} J.,  2023, \mn@doi [arXiv
  e-prints] {10.48550/arXiv.2304.13755}, \href
  {https://ui.adsabs.harvard.edu/abs/2023arXiv230413755M} {p. arXiv:2304.13755}

\bibitem[\protect\citeauthoryear{{McLeod}, {McLure}  \& {Dunlop}}{{McLeod}
  et~al.}{2016}]{McLeod2016}
{McLeod} D.~J.,  {McLure} R.~J.,   {Dunlop} J.~S.,  2016, \mn@doi [\mnras]
  {10.1093/mnras/stw904}, \href
  {https://ui.adsabs.harvard.edu/abs/2016MNRAS.459.3812M} {459, 3812}

\bibitem[\protect\citeauthoryear{{McLeod} et~al.,}{{McLeod}
  et~al.}{2023}]{McLeod2023}
{McLeod} D.~J.,  et~al., 2023, \mn@doi [arXiv e-prints]
  {10.48550/arXiv.2304.14469}, \href
  {https://ui.adsabs.harvard.edu/abs/2023arXiv230414469M} {p. arXiv:2304.14469}

\bibitem[\protect\citeauthoryear{{Meurer}, {Heckman}  \& {Calzetti}}{{Meurer}
  et~al.}{1999}]{Meurer1999}
{Meurer} G.~R.,  {Heckman} T.~M.,   {Calzetti} D.,  1999, \mn@doi [\apj]
  {10.1086/307523}, \href
  {https://ui.adsabs.harvard.edu/abs/1999ApJ...521...64M} {521, 64}

\bibitem[\protect\citeauthoryear{{Mirocha} \& {Furlanetto}}{{Mirocha} \&
  {Furlanetto}}{2023}]{Mirocha2023}
{Mirocha} J.,  {Furlanetto} S.~R.,  2023, \mn@doi [\mnras]
  {10.1093/mnras/stac3578}, \href
  {https://ui.adsabs.harvard.edu/abs/2023MNRAS.519..843M} {519, 843}

\bibitem[\protect\citeauthoryear{{Mirocha}, {La Plante}  \& {Liu}}{{Mirocha}
  et~al.}{2021}]{Mirocha2021}
{Mirocha} J.,  {La Plante} P.,   {Liu} A.,  2021, \mn@doi [\mnras]
  {10.1093/mnras/stab1871}, \href
  {https://ui.adsabs.harvard.edu/abs/2021MNRAS.507.3872M} {507, 3872}

\bibitem[\protect\citeauthoryear{{Morishita} \& {Stiavelli}}{{Morishita} \&
  {Stiavelli}}{2023}]{Morishita2023}
{Morishita} T.,  {Stiavelli} M.,  2023, \mn@doi [\apjl]
  {10.3847/2041-8213/acbf50}, \href
  {https://ui.adsabs.harvard.edu/abs/2023ApJ...946L..35M} {946, L35}

\bibitem[\protect\citeauthoryear{{Morishita} et~al.,}{{Morishita}
  et~al.}{2018}]{Morishita2018}
{Morishita} T.,  et~al., 2018, \mn@doi [\apj] {10.3847/1538-4357/aae68c}, \href
  {https://ui.adsabs.harvard.edu/abs/2018ApJ...867..150M} {867, 150}

\bibitem[\protect\citeauthoryear{{Moster}, {Somerville}, {Maulbetsch}, {van den
  Bosch}, {Macci{\`o}}, {Naab}  \& {Oser}}{{Moster} et~al.}{2010}]{Moster2010}
{Moster} B.~P.,  {Somerville} R.~S.,  {Maulbetsch} C.,  {van den Bosch} F.~C.,
  {Macci{\`o}} A.~V.,  {Naab} T.,   {Oser} L.,  2010, \mn@doi [\apj]
  {10.1088/0004-637X/710/2/903}, \href
  {https://ui.adsabs.harvard.edu/abs/2010ApJ...710..903M} {710, 903}

\bibitem[\protect\citeauthoryear{{Murphy} et~al.,}{{Murphy}
  et~al.}{2011}]{Murphy2011}
{Murphy} E.~J.,  et~al., 2011, \mn@doi [\apj] {10.1088/0004-637X/737/2/67},
  \href {https://ui.adsabs.harvard.edu/abs/2011ApJ...737...67M} {737, 67}

\bibitem[\protect\citeauthoryear{{Murray}}{{Murray}}{2014}]{hmf3}
{Murray} S.,  2014, {HMF: Halo Mass Function calculator}, Astrophysics Source
  Code Library, record ascl:1412.006 (\mn@eprint {ascl} {1412.006})

\bibitem[\protect\citeauthoryear{{Murray}, {Power}  \& {Robotham}}{{Murray}
  et~al.}{2013}]{hmf2}
{Murray} S.~G.,  {Power} C.,   {Robotham} A.~S.~G.,  2013, \mn@doi [Astronomy
  and Computing] {10.1016/j.ascom.2013.11.001}, \href
  {https://ui.adsabs.harvard.edu/abs/2013A&C.....3...23M} {3, 23}

\bibitem[\protect\citeauthoryear{{Naidu} et~al.,}{{Naidu}
  et~al.}{2022a}]{Naidu2022b}
{Naidu} R.~P.,  et~al., 2022a, \mn@doi [arXiv e-prints]
  {10.48550/arXiv.2208.02794}, \href
  {https://ui.adsabs.harvard.edu/abs/2022arXiv220802794N} {p. arXiv:2208.02794}

\bibitem[\protect\citeauthoryear{{Naidu} et~al.,}{{Naidu}
  et~al.}{2022b}]{Naidu2022}
{Naidu} R.~P.,  et~al., 2022b, \mn@doi [\apjl] {10.3847/2041-8213/ac9b22},
  \href {https://ui.adsabs.harvard.edu/abs/2022ApJ...940L..14N} {940, L14}

\bibitem[\protect\citeauthoryear{{Nath}, {Vasiliev}, {Drozdov}  \&
  {Shchekinov}}{{Nath} et~al.}{2023}]{Nath2023}
{Nath} B.~B.,  {Vasiliev} E.~O.,  {Drozdov} S.~A.,   {Shchekinov} Y.~A.,  2023,
  \mn@doi [\mnras] {10.1093/mnras/stad505}, \href
  {https://ui.adsabs.harvard.edu/abs/2023MNRAS.521..662N} {521, 662}

\bibitem[\protect\citeauthoryear{{Oesch}, {Bouwens}, {Illingworth}, {Labb{\'e}}
   \& {Stefanon}}{{Oesch} et~al.}{2018}]{Oesch2018}
{Oesch} P.~A.,  {Bouwens} R.~J.,  {Illingworth} G.~D.,  {Labb{\'e}} I.,
  {Stefanon} M.,  2018, \mn@doi [\apj] {10.3847/1538-4357/aab03f}, \href
  {https://ui.adsabs.harvard.edu/abs/2018ApJ...855..105O} {855, 105}

\bibitem[\protect\citeauthoryear{{Pallottini} \& {Ferrara}}{{Pallottini} \&
  {Ferrara}}{2023}]{Pallottini2023}
{Pallottini} A.,  {Ferrara} A.,  2023, \mn@doi [arXiv e-prints]
  {10.48550/arXiv.2307.03219}, \href
  {https://ui.adsabs.harvard.edu/abs/2023arXiv230703219P} {p. arXiv:2307.03219}

\bibitem[\protect\citeauthoryear{{Papovich} et~al.,}{{Papovich}
  et~al.}{2023}]{Papovich2023}
{Papovich} C.,  et~al., 2023, \mn@doi [\apjl] {10.3847/2041-8213/acc948}, \href
  {https://ui.adsabs.harvard.edu/abs/2023ApJ...949L..18P} {949, L18}

\bibitem[\protect\citeauthoryear{{P{\'e}rez-Gonz{\'a}lez}
  et~al.,}{{P{\'e}rez-Gonz{\'a}lez} et~al.}{2023}]{Perez2023}
{P{\'e}rez-Gonz{\'a}lez} P.~G.,  et~al., 2023, \mn@doi [\apjl]
  {10.3847/2041-8213/acd9d0}, \href
  {https://ui.adsabs.harvard.edu/abs/2023ApJ...951L...1P} {951, L1}

\bibitem[\protect\citeauthoryear{{Planck Collaboration} et~al.,}{{Planck
  Collaboration} et~al.}{2020}]{Planck2020}
{Planck Collaboration} et~al., 2020, \mn@doi [\aap]
  {10.1051/0004-6361/201833910}, \href
  {https://ui.adsabs.harvard.edu/abs/2020A&A...641A...6P} {641, A6}

\bibitem[\protect\citeauthoryear{{Popesso} et~al.,}{{Popesso}
  et~al.}{2023}]{Popesso2023}
{Popesso} P.,  et~al., 2023, \mn@doi [\mnras] {10.1093/mnras/stac3214}, \href
  {https://ui.adsabs.harvard.edu/abs/2023MNRAS.519.1526P} {519, 1526}

\bibitem[\protect\citeauthoryear{{Prada}, {Behroozi}, {Ishiyama}, {Klypin}  \&
  {P{\'e}rez}}{{Prada} et~al.}{2023}]{Prada2023}
{Prada} F.,  {Behroozi} P.,  {Ishiyama} T.,  {Klypin} A.,   {P{\'e}rez} E.,
  2023, \mn@doi [arXiv e-prints] {10.48550/arXiv.2304.11911}, \href
  {https://ui.adsabs.harvard.edu/abs/2023arXiv230411911P} {p. arXiv:2304.11911}

\bibitem[\protect\citeauthoryear{{Press} \& {Schechter}}{{Press} \&
  {Schechter}}{1974}]{Press1974}
{Press} W.~H.,  {Schechter} P.,  1974, \mn@doi [\apj] {10.1086/152650}, \href
  {https://ui.adsabs.harvard.edu/abs/1974ApJ...187..425P} {187, 425}

\bibitem[\protect\citeauthoryear{{Ren}, {Trenti}  \& {Mutch}}{{Ren}
  et~al.}{2018}]{Ren2018}
{Ren} K.,  {Trenti} M.,   {Mutch} S.~J.,  2018, \mn@doi [\apj]
  {10.3847/1538-4357/aab094}, \href
  {https://ui.adsabs.harvard.edu/abs/2018ApJ...856...81R} {856, 81}

\bibitem[\protect\citeauthoryear{Ren, Trenti  \& Mason}{Ren
  et~al.}{2019}]{Ren2019}
Ren K.,  Trenti M.,   Mason C.~A.,  2019, \mn@doi [The Astrophysical Journal]
  {10.3847/1538-4357/ab2117}, 878, 114

\bibitem[\protect\citeauthoryear{{Robertson} et~al.,}{{Robertson}
  et~al.}{2023}]{Robertson2023}
{Robertson} B.~E.,  et~al., 2023, \mn@doi [Nature Astronomy]
  {10.1038/s41550-023-01921-1}, \href
  {https://ui.adsabs.harvard.edu/abs/2023NatAs.tmp...67R} {}

\bibitem[\protect\citeauthoryear{{Rodr{\'\i}guez-Puebla}, {Primack}, {Behroozi}
   \& {Faber}}{{Rodr{\'\i}guez-Puebla} et~al.}{2016}]{Rodrguez2016}
{Rodr{\'\i}guez-Puebla} A.,  {Primack} J.~R.,  {Behroozi} P.,   {Faber} S.~M.,
  2016, \mn@doi [\mnras] {10.1093/mnras/stv2513}, \href
  {https://ui.adsabs.harvard.edu/abs/2016MNRAS.455.2592R} {455, 2592}

\bibitem[\protect\citeauthoryear{{Rogers} et~al.,}{{Rogers}
  et~al.}{2014}]{Rogers2014}
{Rogers} A.~B.,  et~al., 2014, \mn@doi [\mnras] {10.1093/mnras/stu558}, \href
  {https://ui.adsabs.harvard.edu/abs/2014MNRAS.440.3714R} {440, 3714}

\bibitem[\protect\citeauthoryear{{Salpeter}}{{Salpeter}}{1955}]{Salpeter1955}
{Salpeter} E.~E.,  1955, \mn@doi [\apj] {10.1086/145971}, \href
  {https://ui.adsabs.harvard.edu/abs/1955ApJ...121..161S} {121, 161}

\bibitem[\protect\citeauthoryear{{Sheth}, {Mo}  \& {Tormen}}{{Sheth}
  et~al.}{2001}]{Sheth2001}
{Sheth} R.~K.,  {Mo} H.~J.,   {Tormen} G.,  2001, \mn@doi [\mnras]
  {10.1046/j.1365-8711.2001.04006.x}, \href
  {https://ui.adsabs.harvard.edu/abs/2001MNRAS.323....1S} {323, 1}

\bibitem[\protect\citeauthoryear{{Smit}, {Bouwens}, {Labb{\'e}}, {Franx},
  {Wilkins}  \& {Oesch}}{{Smit} et~al.}{2016}]{Smit2016}
{Smit} R.,  {Bouwens} R.~J.,  {Labb{\'e}} I.,  {Franx} M.,  {Wilkins} S.~M.,
  {Oesch} P.~A.,  2016, \mn@doi [\apj] {10.3847/1538-4357/833/2/254}, \href
  {https://ui.adsabs.harvard.edu/abs/2016ApJ...833..254S} {833, 254}

\bibitem[\protect\citeauthoryear{{Sparre}, {Hayward}, {Feldmann},
  {Faucher-Gigu{\`e}re}, {Muratov}, {Kere{\v s}}  \& {Hopkins}}{{Sparre}
  et~al.}{2017}]{Sparre2015}
{Sparre} M.,  {Hayward} C.~C.,  {Feldmann} R.,  {Faucher-Gigu{\`e}re} C.-A.,
  {Muratov} A.~L.,  {Kere{\v s}} D.,   {Hopkins} P.~F.,  2017, \mn@doi [\mnras]
  {10.1093/mnras/stw3011}, \href
  {http://adsabs.harvard.edu/abs/2017MNRAS.466...88S} {466, 88}

\bibitem[\protect\citeauthoryear{{Speagle}, {Steinhardt}, {Capak}  \&
  {Silverman}}{{Speagle} et~al.}{2014}]{Speagle2014}
{Speagle} J.~S.,  {Steinhardt} C.~L.,  {Capak} P.~L.,   {Silverman} J.~D.,
  2014, \mn@doi [\apjs] {10.1088/0067-0049/214/2/15}, \href
  {https://ui.adsabs.harvard.edu/abs/2014ApJS..214...15S} {214, 15}

\bibitem[\protect\citeauthoryear{{Springel} et~al.,}{{Springel}
  et~al.}{2005}]{Springel2005}
{Springel} V.,  et~al., 2005, \mn@doi [\nat] {10.1038/nature03597}, \href
  {https://ui.adsabs.harvard.edu/abs/2005Natur.435..629S} {435, 629}

\bibitem[\protect\citeauthoryear{{Stefanon} et~al.,}{{Stefanon}
  et~al.}{2019}]{Stefanon2019}
{Stefanon} M.,  et~al., 2019, \mn@doi [\apj] {10.3847/1538-4357/ab3792}, \href
  {https://ui.adsabs.harvard.edu/abs/2019ApJ...883...99S} {883, 99}

\bibitem[\protect\citeauthoryear{{Steinhardt}, {Kokorev}, {Rusakov}, {Garcia}
  \& {Sneppen}}{{Steinhardt} et~al.}{2023}]{Steinhardt2022}
{Steinhardt} C.~L.,  {Kokorev} V.,  {Rusakov} V.,  {Garcia} E.,   {Sneppen} A.,
   2023, \mn@doi [\apjl] {10.3847/2041-8213/acdef6}, \href
  {https://ui.adsabs.harvard.edu/abs/2023ApJ...951L..40S} {951, L40}

\bibitem[\protect\citeauthoryear{{Sun} \& {Furlanetto}}{{Sun} \&
  {Furlanetto}}{2016}]{Sun2016}
{Sun} G.,  {Furlanetto} S.~R.,  2016, \mn@doi [\mnras] {10.1093/mnras/stw980},
  \href {https://ui.adsabs.harvard.edu/abs/2016MNRAS.460..417S} {460, 417}

\bibitem[\protect\citeauthoryear{{Sun}, {Faucher-Gigu{\`e}re}, {Hayward},
  {Shen}, {Wetzel}  \& {Cochrane}}{{Sun} et~al.}{2023}]{Sun2023}
{Sun} G.,  {Faucher-Gigu{\`e}re} C.-A.,  {Hayward} C.~C.,  {Shen} X.,  {Wetzel}
  A.,   {Cochrane} R.~K.,  2023, \mn@doi [arXiv e-prints]
  {10.48550/arXiv.2307.15305}, \href
  {https://ui.adsabs.harvard.edu/abs/2023arXiv230715305S} {p. arXiv:2307.15305}

\bibitem[\protect\citeauthoryear{{Tacchella}, {Trenti}  \&
  {Carollo}}{{Tacchella} et~al.}{2013}]{Tacchella2013}
{Tacchella} S.,  {Trenti} M.,   {Carollo} C.~M.,  2013, \mn@doi [\apjl]
  {10.1088/2041-8205/768/2/L37}, \href
  {https://ui.adsabs.harvard.edu/abs/2013ApJ...768L..37T} {768, L37}

\bibitem[\protect\citeauthoryear{{Tacchella}, {Dekel}, {Carollo}, {Ceverino},
  {DeGraf}, {Lapiner}, {Mandelker}  \& {Primack Joel}}{{Tacchella}
  et~al.}{2016}]{Tacchella2016}
{Tacchella} S.,  {Dekel} A.,  {Carollo} C.~M.,  {Ceverino} D.,  {DeGraf} C.,
  {Lapiner} S.,  {Mandelker} N.,   {Primack Joel} R.,  2016, \mn@doi [\mnras]
  {10.1093/mnras/stw131}, \href
  {https://ui.adsabs.harvard.edu/abs/2016MNRAS.457.2790T} {457, 2790}

\bibitem[\protect\citeauthoryear{Tacchella, Bose, Conroy, Eisenstein  \&
  Johnson}{Tacchella et~al.}{2018}]{Tacchella2018}
Tacchella S.,  Bose S.,  Conroy C.,  Eisenstein D.~J.,   Johnson B.~D.,  2018,
  \mn@doi [The Astrophysical Journal] {10.3847/1538-4357/aae8e0}, 868, 92

\bibitem[\protect\citeauthoryear{{Tacchella}, {Forbes}  \&
  {Caplar}}{{Tacchella} et~al.}{2020}]{Tacchella2020}
{Tacchella} S.,  {Forbes} J.~C.,   {Caplar} N.,  2020, \mn@doi [\mnras]
  {10.1093/mnras/staa1838}, \href
  {https://ui.adsabs.harvard.edu/abs/2020MNRAS.497..698T} {497, 698}

\bibitem[\protect\citeauthoryear{{Tan}}{{Tan}}{2000}]{Tan2000}
{Tan} J.~C.,  2000, \mn@doi [\apj] {10.1086/308905}, \href
  {https://ui.adsabs.harvard.edu/abs/2000ApJ...536..173T} {536, 173}

\bibitem[\protect\citeauthoryear{{Tasker}}{{Tasker}}{2011}]{Tasker2011}
{Tasker} E.~J.,  2011, \mn@doi [\apj] {10.1088/0004-637X/730/1/11}, \href
  {https://ui.adsabs.harvard.edu/abs/2011ApJ...730...11T} {730, 11}

\bibitem[\protect\citeauthoryear{{Tinker}, {Robertson}, {Kravtsov}, {Klypin},
  {Warren}, {Yepes}  \& {Gottl{\"o}ber}}{{Tinker} et~al.}{2010}]{Tinker2010}
{Tinker} J.~L.,  {Robertson} B.~E.,  {Kravtsov} A.~V.,  {Klypin} A.,  {Warren}
  M.~S.,  {Yepes} G.,   {Gottl{\"o}ber} S.,  2010, \mn@doi [\apj]
  {10.1088/0004-637X/724/2/878}, \href
  {https://ui.adsabs.harvard.edu/abs/2010ApJ...724..878T} {724, 878}

\bibitem[\protect\citeauthoryear{{Topping}, {Stark}, {Endsley}, {Plat},
  {Whitler}, {Chen}  \& {Charlot}}{{Topping} et~al.}{2022}]{Topping2022}
{Topping} M.~W.,  {Stark} D.~P.,  {Endsley} R.,  {Plat} A.,  {Whitler} L.,
  {Chen} Z.,   {Charlot} S.,  2022, \mn@doi [\apj] {10.3847/1538-4357/aca522},
  \href {https://ui.adsabs.harvard.edu/abs/2022ApJ...941..153T} {941, 153}

\bibitem[\protect\citeauthoryear{{Trenti} \& {Stiavelli}}{{Trenti} \&
  {Stiavelli}}{2008}]{Trenti2008}
{Trenti} M.,  {Stiavelli} M.,  2008, \mn@doi [\apj] {10.1086/528674}, \href
  {https://ui.adsabs.harvard.edu/abs/2008ApJ...676..767T} {676, 767}

\bibitem[\protect\citeauthoryear{{Treu} et~al.,}{{Treu}
  et~al.}{2023}]{Treu2023}
{Treu} T.,  et~al., 2023, \mn@doi [\apjl] {10.3847/2041-8213/ac9283}, \href
  {https://ui.adsabs.harvard.edu/abs/2023ApJ...942L..28T} {942, L28}

\bibitem[\protect\citeauthoryear{{Vogelsberger} et~al.,}{{Vogelsberger}
  et~al.}{2020}]{Vogelsberger2020}
{Vogelsberger} M.,  et~al., 2020, \mn@doi [\mnras] {10.1093/mnras/staa137},
  \href {https://ui.adsabs.harvard.edu/abs/2020MNRAS.492.5167V} {492, 5167}

\bibitem[\protect\citeauthoryear{{Whitler}, {Mason}, {Ren}, {Dijkstra},
  {Mesinger}, {Pentericci}, {Trenti}  \& {Treu}}{{Whitler}
  et~al.}{2020}]{Whitler2020}
{Whitler} L.~R.,  {Mason} C.~A.,  {Ren} K.,  {Dijkstra} M.,  {Mesinger} A.,
  {Pentericci} L.,  {Trenti} M.,   {Treu} T.,  2020, \mn@doi [\mnras]
  {10.1093/mnras/staa1178}, \href
  {https://ui.adsabs.harvard.edu/abs/2020MNRAS.495.3602W} {495, 3602}

\bibitem[\protect\citeauthoryear{{Wilkins} et~al.,}{{Wilkins}
  et~al.}{2023a}]{Wilkins2023b}
{Wilkins} S.~M.,  et~al., 2023a, \mn@doi [\mnras] {10.1093/mnras/stac3281},
  \href {https://ui.adsabs.harvard.edu/abs/2023MNRAS.518.3935W} {518, 3935}

\bibitem[\protect\citeauthoryear{{Wilkins} et~al.,}{{Wilkins}
  et~al.}{2023b}]{Wilkins2023}
{Wilkins} S.~M.,  et~al., 2023b, \mn@doi [\mnras] {10.1093/mnras/stac3280},
  \href {https://ui.adsabs.harvard.edu/abs/2023MNRAS.519.3118W} {519, 3118}

\bibitem[\protect\citeauthoryear{{Yan}, {Ma}, {Ling}, {Cheng}  \&
  {Huang}}{{Yan} et~al.}{2023}]{Yan2023}
{Yan} H.,  {Ma} Z.,  {Ling} C.,  {Cheng} C.,   {Huang} J.-S.,  2023, \mn@doi
  [\apjl] {10.3847/2041-8213/aca80c}, \href
  {https://ui.adsabs.harvard.edu/abs/2023ApJ...942L...9Y} {942, L9}

\bibitem[\protect\citeauthoryear{{Yung}, {Somerville}, {Finkelstein}, {Popping}
   \& {Dav{\'e}}}{{Yung} et~al.}{2019}]{Yung2019}
{Yung} L.~Y.~A.,  {Somerville} R.~S.,  {Finkelstein} S.~L.,  {Popping} G.,
  {Dav{\'e}} R.,  2019, \mn@doi [\mnras] {10.1093/mnras/sty3241}, \href
  {https://ui.adsabs.harvard.edu/abs/2019MNRAS.483.2983Y} {483, 2983}

\bibitem[\protect\citeauthoryear{{Yung}, {Somerville}, {Finkelstein}, {Wilkins}
   \& {Gardner}}{{Yung} et~al.}{2023}]{Yung2023}
{Yung} L.~Y.~A.,  {Somerville} R.~S.,  {Finkelstein} S.~L.,  {Wilkins} S.~M.,
  {Gardner} J.~P.,  2023, \mn@doi [arXiv e-prints] {10.48550/arXiv.2304.04348},
  \href {https://ui.adsabs.harvard.edu/abs/2023arXiv230404348Y} {p.
  arXiv:2304.04348}

\bibitem[\protect\citeauthoryear{{Zavala} et~al.,}{{Zavala}
  et~al.}{2023}]{Zavala2023}
{Zavala} J.~A.,  et~al., 2023, \mn@doi [\apjl] {10.3847/2041-8213/acacfe},
  \href {https://ui.adsabs.harvard.edu/abs/2023ApJ...943L...9Z} {943, L9}

\bibitem[\protect\citeauthoryear{{Ziparo}, {Ferrara}, {Sommovigo}  \&
  {Kohandel}}{{Ziparo} et~al.}{2023}]{Ziparo2023}
{Ziparo} F.,  {Ferrara} A.,  {Sommovigo} L.,   {Kohandel} M.,  2023, \mn@doi
  [\mnras] {10.1093/mnras/stad125}, \href
  {https://ui.adsabs.harvard.edu/abs/2023MNRAS.520.2445Z} {520, 2445}

\makeatother
\end{thebibliography}






\bsp	
\label{lastpage}
\end{document}